\pdfoutput=1
\documentclass[useAMS,usenatbib]{mn2e}
\usepackage{amssymb}
\usepackage{bbm}
\usepackage{txfonts}
\usepackage{psfrag}
\usepackage{graphicx}
\usepackage{natbib}
\usepackage{if then}
\usepackage{longtable}
\usepackage{color}
\usepackage{tikz}
\usetikzlibrary{shapes,arrows}
\usepackage{verbatim}

\def\del#1{{}}

\newcommand{\unit}[1]{\mbox{  }\rm{#1}}

\title{Bayesian inference from photometric redshift surveys}
\author[Jens Jasche \\$^{1}$, Benjamin D. Wandelt]
       {Jens Jasche $^{1}$,Benjamin D. Wandelt $^{2,3,4}$ \\$^{1}$ Argelander-Institut f\"{u}r Astronomie , Auf dem H\"{u}gel 71,  D-53121 Bonn, Germany\\$^{2}$ UPMC Univ Paris 06, UMR 7095, Institut d'Astrophysique de Paris, 98 bis, boulevard Arago, 75014 Paris, France\\$^{3}$ CNRS, UMR 7095, Institut d'Astrophysique de Paris, 98 bis, boulevard Arago, 75014 Paris, France\\$^{4}$ Departments of Physics and Astronomy, 1110 W Green Street, University of Illinois at Urbana-Champaign, Urbana, IL 61801, USA}

\begin{document}

\date{Submitted to MNRAS 14.06.2011}

\pagerange{\pageref{firstpage}--\pageref{lastpage}} \pubyear{2006}

\maketitle

\label{firstpage}

\begin{abstract}
We show how to enhance the redshift accuracy of surveys  consisting of tracers with highly uncertain positions along the line of sight. Photometric surveys with redshift uncertainty $\delta z\sim 0.03$ can yield final redshift uncertainties of $\delta z_{f}\sim 0.003$ in high density regions. This increased redshift precision is achieved by imposing an isotropy and 2-point correlation prior in a Bayesian analysis and is completely independent of the process that estimates the photometric redshift. As a byproduct, the method also infers the three dimensional density field, essentially super-resolving high density regions in redshift space. Our method fully takes into account the survey mask and selection function. It uses a simplified Poissonian picture of galaxy formation, relating preferred locations of galaxies to regions of higher density in the matter field. The method quantifies the remaining uncertainties in the three dimensional density field and the true radial locations of galaxies by generating samples that are constrained by the survey data. The exploration of this high dimensional, non-Gaussian joint posterior is made feasible using multiple-block Metropolis-Hastings sampling. We demonstrate the performance of our implementation on a simulation containing $2\times 10^{7}$ galaxies. These results bear out the promise of Bayesian analysis  for upcoming photometric large scale structure surveys with tens of millions of galaxies.
\end{abstract}


\section{Introduction}
\label{INTRO}
Understanding the origin and evolution of the Universe to high precision requires increasing amounts of data. In particular, understanding the nature of the recently observed accelerated expansion, measuring the Baryon acoustic scale in the matter distribution or establishing new standard rulers require probes of the three dimensional galaxy distribution over large volumes, out to high redshifts. A particularly difficult problem in galaxy observations arises from the requirement of accurate redshift measurements as input to high precision cosmology. Spectroscopy is the most accurate method for measuring the radial distances of galaxies along the line of sight. However, it is time-intensive and needs high photon fluxes. Therefore spectroscopic redshifts will be unavailable for the vast majority of the (tens of) millions of galaxies to be observed by future galaxy surveys \citep[][]{BENJAMIN2010}. 
These requirements have paved the way for photometric redshift (photo-z) measurements as valuable astronomical tools in future observations of the galaxy distribution. Photometric galaxy observations can not only measure redshifts for fainter objects than could be yielded with spectroscopy but also provide higher numbers of objects with redshift measurements per unit telescope time \citep[][]{Hildebrandt2010}.
Consequently, using photometry rather than spectroscopy permits exploring the galaxy distribution to higher redshifts at the expense of increased uncertainty in the determination of three dimensional galaxy positions \citep{BAUM1962,Koo1986,Connolly1995,FERNANDEZSOTO2001,BLAKE2005}. 

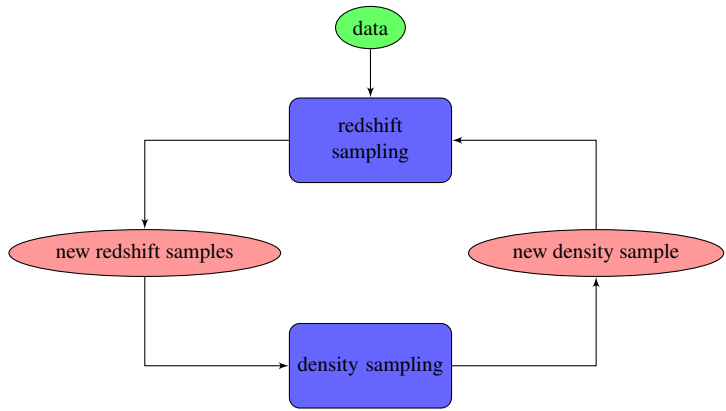
\begin{figure*}
	\centering
	{
	\tikzstyle{blank} = [rectangle,  fill=white!20,text width=5em, text centered, rounded corners, minimum height=4em]
	\tikzstyle{decision} = [diamond, draw, fill=blue!40,text width=4.5em, text badly centered, node distance=3cm, inner sep=0pt]
	\tikzstyle{block} = [rectangle, draw, fill=blue!60,text width=7em, text centered, rounded corners, minimum height=4em]
	\tikzstyle{line} = [draw, -latex']
	\tikzstyle{cloud} = [draw, ellipse,fill=red!40, node distance=3cm, minimum height=2em]
	\tikzstyle{clouda} = [draw, ellipse,fill=green!60, node distance=3cm, minimum height=2em]
\begin{tikzpicture}[node distance = 1.5cm, auto]
    \node [clouda] (data) {data};
    \node [block,below of=data] (DENSSAMPLING) {redshift sampling};
    \node [blank,below of=DENSSAMPLING] (blank) {};
    \node [block,below of=blank] (ZSAMPLING) {density sampling};	
    \node [cloud,left of=blank] (denssample) {new redshift samples};
    \node [cloud,right of=blank] (zsample) {new density sample};	
    \path [line] (data) -- (DENSSAMPLING);
    \path [line] (DENSSAMPLING) -| (denssample);
    \path [line] (denssample) |- (ZSAMPLING);	
    \path [line] (ZSAMPLING) -| (zsample);
    \path [line] (zsample) |- (DENSSAMPLING);
\end{tikzpicture}
	}
\caption{Flow-chart depicting the two step iterative block  sampling procedure. First a set of galaxy redshift reconstructions is sampled conditional on the last density field. In a second step, a density field sample is generated conditional on the new galaxy redshifts, assuming them to be related to the density field through log-normal Poisson statistics.}
	\label{fig:flowchart}
\end{figure*}

Since major on-going and future galaxy surveys, such as DES, PAU, CFHTLS and the space-based EUCLID mission, will be photometric, redshift uncertainties will play an increasingly important role in future observations of the large scale structure \citep[see e.g.][]{COUPON2009,BENITEZ2009,REFREGIER2010,FRIEMAN2011}. Nevertheless, precision cosmology requires accurate distance estimates. The impact of redshift uncertainties on the inference of cosmological signals such as the baryon acoustic oscillations, cluster counts or weak lensing has been extensively discussed in the literature \citep[see e.g.][]{HUTERER2004,BERNSTEIN2004,ISHAK2005,HUTERER2006,MA2006,ZHAN2006,ZHAN2006B}.

In this work we present a new Bayesian approach for redshift inference from galaxy redshift surveys with uncertain radial locations of galaxies. We emphasize that our method is independent and complementary to the methods inferring photo-zs from color or morphology information, and is therefore intended to be used in conjunction with those.
In order to increase the accuracy of galaxy positions our method exploits two physically motivated assumptions. The first assumption relies on the cosmological principle which states that the Universe should be homogeneous and isotropic. For this reason, the covariance matrix of the matter distribution in Fourier space is of diagonal shape, at least at the largest scales, with the diagonal being the cosmological power-spectrum. As a consequence, the sampler searches for galaxy redshift configurations which are compatible with an isotropic matter distribution. Secondly, we exploit a simplified physical picture of galaxy formation, which states that galaxies predominantly form in regions of higher density.

In practice neither the three dimensional matter field nor the exact galaxy positions are known  \textit{a priori}. We must infer the three dimensional density field together with the  locations of galaxies. In so doing, we account for the mutual dependences and correlated uncertainties of the inferred density field and galaxy positions, and show that the joint treatment can enhance their mutual inference.

In order to perform a full joint Bayesian analysis of the three dimensional density field and the  redshifts of galaxies we implement a two step Markov sampling process as depicted in figure \ref{fig:flowchart}, which permits to explore the high dimensional non-Gaussian joint posterior distribution in a numerically efficient way.
In particular, within the framework of a multiple block Metropolis-Hastings sampler, we are able to break up the high dimensional problem in a set of lower dimensional problems which can then be treated iteratively \citep[][]{hastings}.
This  sampling approach permits us to produce data-constrained samples of the three dimensional density field and galaxy redshifts for photo-z surveys with tens of millions of galaxies in a numerically efficient way.

In this work we are particularly interested in the final redshift uncertainties, which are output as sampled probability densities of galaxy redshifts constrained by the observations and incorporating the isotropy prior. These results improve estimates for the radial locations of galaxies and  quantify the remaining uncertainties.

The paper is structured as follows. In section \ref{method} we give a brief overview and introduction to the method presented in this work. A brief overview of photo-z estimators which generate the input needed for our work is provided in appendix \ref{photozmethods}. The density field sampling step and the corresponding log-normal Poissonian model for the density field is briefly discussed in section \ref{log_normal_model}. Since this sampling step has already been solved within the HADES (Hamiltonian Density Estimation and Sampling) algorithm a more detailed description can be found in \citet{JASCHE2010HADESMETHOD} and \citet{JASCHE2010HADESDATA}. 
The redshift sampling process and the derivation of the corresponding conditional redshift posterior distribution is discussed in section \ref{red_post_dist}. In section \ref{mock_observations} we describe the generation of the mock data used in our tests.

\begin{figure*}
\centering{\includegraphics{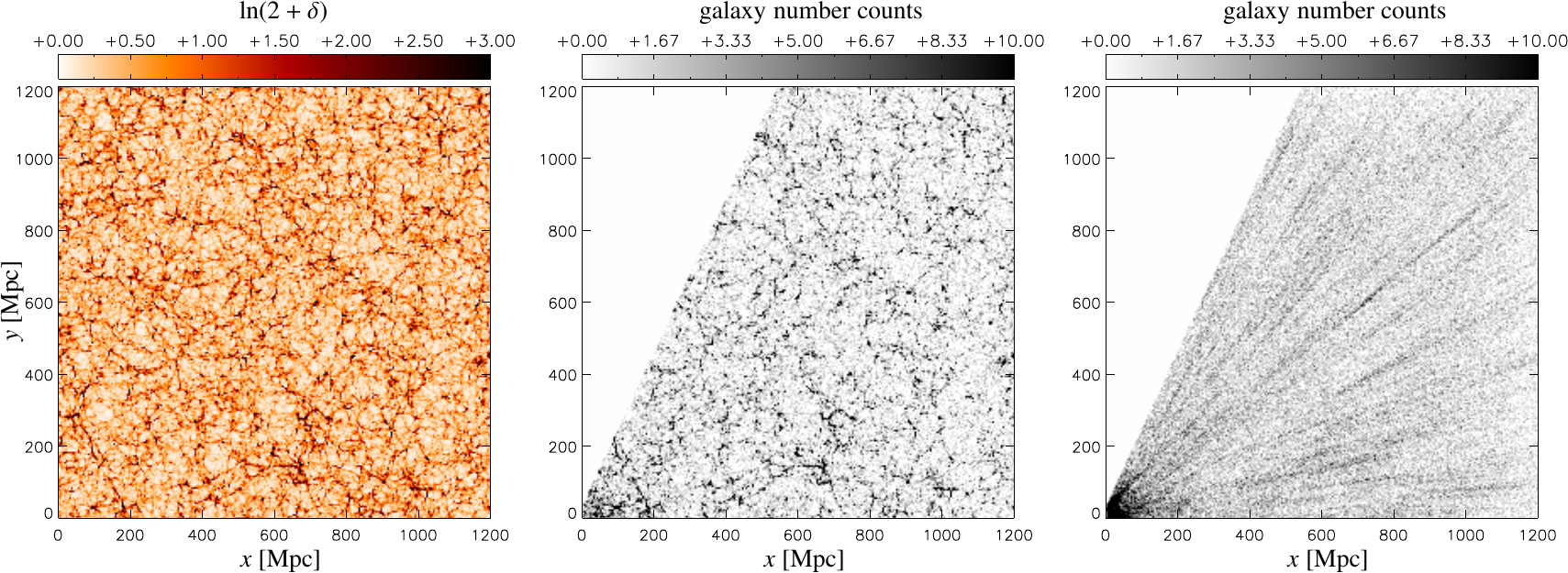}}
\caption{Slices through the mock matter field realization (left panel), the mock galaxy survey generated from the mock matter field without distorted redshifts (middle panel) and the redshift distorted mock galaxy survey (right panel). It is clearly visible that all small scale structures are smeared out along the line of sight in a survey with photo-z errors of $\sim 0.03$.}
\label{fig:mock}
\end{figure*}

The performance of our multi dimensional Metropolis Hastings sampler in a realistic setting is studied in section \ref{TESTING}. Here we study the overall burn-in and convergence behavior of the Markov sampler.  In section \ref{results} we discuss the results for the inferred redshifts and three dimensional density fields. Finally, in section \ref{discussion}, we discuss our results and give an outlook on future improvements and applications of our work.

\section{Method}
\label{method}
The method presented in this work aims at the joint inference of radial locations of galaxies as well as three dimensional density fields from observations with uncertain redshift measurements. In particular, we are interested in exploring the joint probability \({\cal P}(\{s_i\},\{z_p\}|\{z^{obs}_n\},\{\theta_n\})\) of the real-space density field \(\{s_i\}\) and the true redshift positions of galaxies \(\{z_p\}\) conditional on the observed redshifts \(\{z^{obs}_n\}\) and the positions in the sky \(\{\theta_n\}\), which is a set containing the declination and right ascension angles for all galaxies. The exploration of the joint probability distribution is numerically challenging partly due to high dimensionality but also because the problem requires non-linear and non-Gaussian inference methods. In situations like this, one usually has to rely on efficient implementations of a Metropolis Hastings sampling scheme. However, it may be difficult to construct a single block Markov algorithm which converges rapidly to the target distribution. 

Fortunately, it is possible to break up the parameter space into smaller blocks and to construct transition kernels for each individual block \citep[][]{hastings}. Consequently this multiple block sampling framework permits to to build complex transition kernels for the joint problem. This approach greatly simplifies the search for suitable candidate-generating densities and enables us to use the previously developed non-linear log-normal Poissonian density sampler, presented in  \citep[][]{JASCHE2010HADESMETHOD,JASCHE2010HADESDATA}, in conjunction with a new redshift sampler described in this work. This block sampling approach was also at the heart of the Gibbs sampler for cosmic microwave background  (CMB) power spectrum inference originally proposed in \citet[][]{WANDELT2004} and \citet[][]{2004ApJ...609....1J} which is now widely used.

We can explore the joint posterior distribution by iteratively sampling from the two conditional distributions
\begin{eqnarray}
\label{eq:MARKOV_STEPS}
& {\rm{1}}) & \{z_p\}^{(j+1)}\curvearrowleft {\cal P}(\{z_p\}|\{s_i\}^{(j)},\{z^{obs}_p\},\{\theta_p\})\, , \nonumber \\
& {\rm{2}}) & \{s_i\}^{(j+1)}\curvearrowleft {\cal P}(\{s_i\}|\{z_p\}^{(j+1)},\{z^{obs}_p\},\{\theta_p\}) \, , 
\end{eqnarray}
where in the first step we draw a random realization of radial galaxy positions \(\{z_p\}\) and in a second step we generate a three dimensional density field sample \(\{s_i\}\). The iteration of these two random processes will provide us with samples of the joint posterior distribution. A schematic description of the sampling procedure is depicted in figure \ref{fig:flowchart}.

\begin{figure*}
\centering{\includegraphics{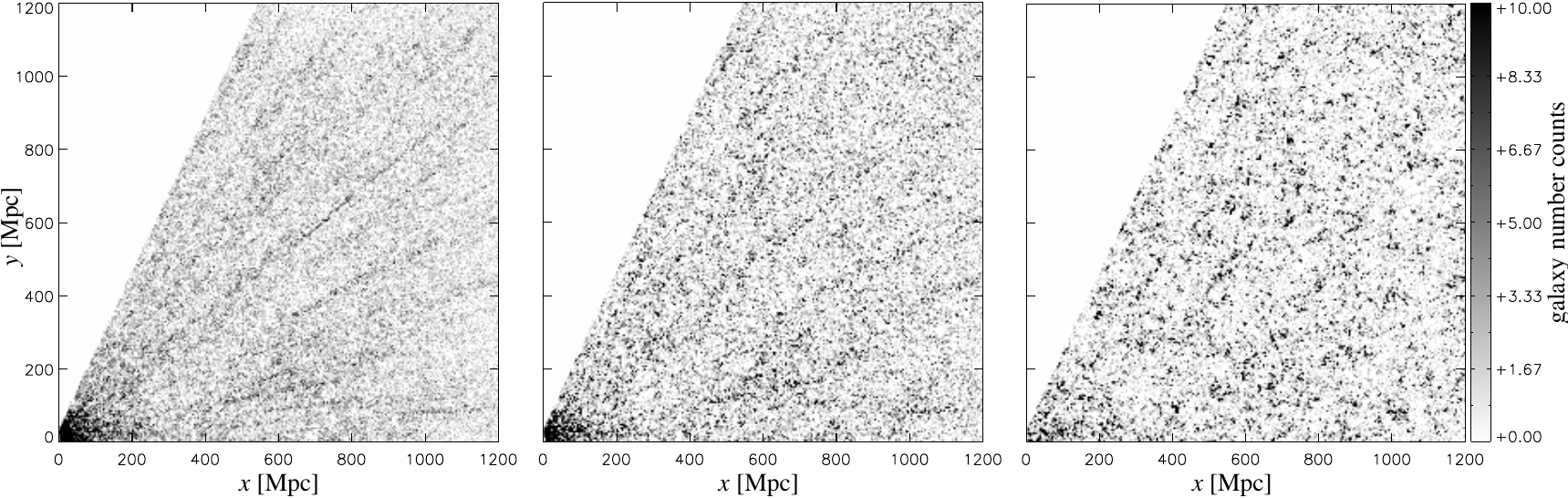}}
\caption{Slices through three successive galaxy count samples, first sample (left panel), fifth sample middle panel and 100th sample (right panel). The first sample shows prominent redshift smearing which quickly decay in subsequent samples.}
\label{fig:burn_in_z}
\end{figure*}

\section{The log normal Poissonian model}
\label{log_normal_model} 
 The log-normal Poissonian posterior distribution is formulated in terms of galaxy number counts, rather than in terms of galaxy redshifts and positions in the sky.
As demonstrated in appendix \ref{Appendix_densitypost}, the density posterior distribution given in equation (\ref{eq:MARKOV_STEPS}) can be written as a function of galaxy number counts
\begin{eqnarray}
{\cal P}(\{s_i\}|\{z_p\},\{z^{obs}_p\},\{\theta_p\}) &=& {\cal P}\left(\{s_i\}|\left \{N_i\left(\{z_p\},\{\theta_p\}\right)\right \} \right) \, ,
\end{eqnarray}
where the number counts \(N_i\left(\{z_p\},\{\theta_p\}\right) \) are calculated according to equation (\ref{eq:appendix_calc_ncounts}).
Following \citet{JASCHE2010HADESMETHOD}, the log normal Poissonian posterior distribution can then be expressed as:

\begin{eqnarray}
\label{eq:LOGNORMALPOISSONIAN_POSTERIOR}
{\mathcal P}(\{s_i\}|\{N_i\}) &=& \frac{e^{-\frac{1}{2}\sum_{ij} \left(\mathrm{ln}(1+s_i)+\mu_i\right) Q^{-1}_{ij} \left(\mathrm{ln}(1+s_j)+\mu_j\right)}}{\sqrt{|2\pi Q|}} \prod_l \frac{1}{1+s_l} \nonumber \\
& & \times \prod_k \frac{{\left(R_k \bar{N}(1+B(s)_k)\right)}^{N^{g}_k} e^{-R_k \bar{N}(1+B(s)_k)}}{{N^{g}_k}!}\, ,
\end{eqnarray}
where \(Q\) is the covariance matrix of the log-normal distribution, \(\mu_i\) describes a constant mean field, \(R_k\) is a linear response operator, incorporating survey geometries and selection effects, \(\bar{N}\) is the mean number of galaxies in the volume and \(B(x)_k\) is a non-linear, non local, bias operator at position \(\vec{x}_k\).
In this fashion the log-normal Poissonian distribution accounts for non-linearities in the density field, the local noise structure of the galaxy distribution as well as systematic uncertainties due to survey geometry, selection effects or galaxy biasing.

The log-normal posterior distribution incorporates essentially two physically motivated assumptions which are important for the method described in this work. Due to the cosmological principle we will assume the matter distribution in real-space to be statistically isotropic. This is incorporated in the posterior distribution in equation (\ref{eq:LOGNORMALPOISSONIAN_POSTERIOR}) by noting, that the Fourier transform of the covariance matrix \(Q\) for the logarithmic density field is assumed to have a diagonal form with the diagonal being the power-spectrum which depends on the $|k|$ only. This assumption requires the individual density field samples to be statistically isotropic. In addition the Poisson model can be considered as a simplified model for galaxy formation, which relates the number of galaxies found at a given position to the underlying density field, and hence singles out preferred locations for galaxies conditional on the underlying density field. Thus, in the absence of additional constraining information, these physical requirements enhance the inferred redshifts from observations with uncertain radial locations of galaxies.  

For a detailed description on the numerical implementation of the large scale structure sampler we refer to \citet[][]{JASCHE2010HADESMETHOD}. 
Unlike conventional Metropolis Hastings algorithms, which move through the parameter space by a random walk, and therefore require prohibitive amounts of steps to explore high dimensional spaces, the HMC sampler suppresses random walk behavior by introducing persistent motions of the Markov chain through the parameter space \citep[][]{DUANE1987,NEAL1993,NEAL1996}. In this fashion, the Hamiltonian sampler maintains a reasonable efficiency even for high dimensional problems \citep{HANSON2001}.

\section{The redshift posterior distribution}
\label{red_post_dist}

In the following we will focus on the derivation and implementation of the second sampling step in figure \ref{fig:flowchart}, namely of generating random realizations of true galaxy redshifts \(\{z_p\}\) conditional on observations and a realization of the three dimensional density field. In particular we are searching for a numerically efficient implementation of the random process
\begin{equation}
\label{eq:joint_redshift posterior}
\{z_p\}\curvearrowleft {\cal P}(\{z_p\}|\{s_i\},\{z^{obs}_p\},\{\theta_p\})\, . 
\end{equation}
This problem is numerically challenging due to the high dimensionality, i.e. the number of galaxies in a survey. Additionally, the non-linearity of the problem generally prevents   drawing samples directly from the redshift posterior distribution given in equation (\ref{eq:joint_redshift posterior}). Hence, in order to explore the space of possible galaxy redshift configurations we have to rely on a Metropolis-Hastings sampler.
A naive implementation of the redshift sampling process might suggest drawing a realization for the whole set \(\{z_p\}\) at once. Such a scheme will  lead to a high rejection rate, rendering such a method numerically unfeasible. Suppose, for example, that the sampler proposes a redshift configuration \(\{z_p\}\) which is acceptable except for one galaxy redshift \(z_n\). Although most of the galaxy redshifts could be accepted in this scenario the complete set \(\{z_p\}\) would be rejected and the sampler stays at its current position in parameter space.   For this reason, such a sampling approach would require to construct a very good proposal distribution for \( {\cal P}(\{z_p\}|\{s_i\},\{z^{obs}_p\},\{\theta_p\})\) in order to yield efficient sampling rates. Generally it is not trivial and may even be impossible to construct such a proposal distribution.

If it were possible to find conditional redshift posterior distributions for each individual galaxy redshift \(z_n\), one could build a multiple block Metropolis-Hastings sampler for the set of galaxy redshifts \(\{z_p\}\). 

Fortunately, this is possible in our simple, Poissonian galaxy formation model. As demonstrated in appendix \ref{Appendix_densitypost} it is possible to find a conditional redshift posterior distribution \({\cal P}(z_n|\{u_p\},\{s_i\},\{z^{obs}_p\},\{\theta_p\})\) such that a multiple block Metropolis-Hastings sampler can be constructed. 

Galaxy redshifts can therefore be updated by sampling \( {\cal P}(\{z_p\}|\{s_i\},\{z^{obs}_p\},\{\theta_p\})\) using the following random process
\begin{eqnarray}
\label{eq:multiblock}
& {\rm{1}}) & z_1^{(j+1)}\curvearrowleft {\cal P}(z_1|\{u_p\}^1,\{s_i\},\{z^{obs}_n\},\{\theta_n\}) \nonumber \\
& {\rm{2}}) & z_2^{(j+1)}\curvearrowleft {\cal P}(z_2|\{u_p\}^2,\{s_i\},\{z^{obs}_n\},\{\theta_n\}) \nonumber \\ 
& {\rm{3}}) & z_3^{(j+1)}\curvearrowleft {\cal P}(z_3|\{u_p\}^3,\{s_i\},\{z^{obs}_n\},\{\theta_n\}) \nonumber \\
& {\rm{.}} & \nonumber \\
& {\rm{.}} & \nonumber \\
& {\rm{.}} & \nonumber \\
& {\rm{N}}) & z_N^{(j+1)}\curvearrowleft {\cal P}(z_N|\{u_p\}^N,\{s_i\},\{z^{obs}_n\},\{\theta_n\})\, , 
\end{eqnarray}
where for each transition we use a Metropolis-Hastings update step and \(\{u_p\}=\{z_p\}_{/z_n}\) is the set of galaxy redshifts containing all redshifts except the redshift \(z_n\) of the \(n\)th galaxy under consideration. The superscripts \(i\) at the \(\{u_p\}^i\) indicate that the set of galaxy redshifts \(\{u_p\}\) have to be updated after each sampling step, since the next sampling step depends on the new position of the previously sampled galaxy.
The conditional redshift posteriors \({\cal P}(z_n|\{u_p\},\{s_i\},\{z^{obs}_n\},\{\theta_n\})\)  describes the probability for the redshift \(z_n\) of the \(n\)th galaxy redshift conditional on the redshifts of all other galaxies, the density field, the observed redshifts and the positions in the sky.

At this point an important further simplification occurs.  Given the density field $s$, the redshifts any one galaxy \(z_n\) is conditionally independent of the redshifts of \textit{ all the other galaxies}, so
\begin{equation}
{\cal P}(z_n|\{u_p\},\{s_i\},\{z^{obs}_p\},\{\theta_p\})={\cal P}(z_n|\{s_i\},\{z^{obs}_p\},\{\theta_p\})\, ,
\end{equation}
which is also shown in appendix \ref{Appendix_densitypost}. This is particularly interesting for the numerical implementation of the random process given in equation (\ref{eq:multiblock}), since it permits parallel, instead of sequential, sampling.

\begin{figure*}
\centering{\includegraphics[width=1.0\textwidth,clip=true]{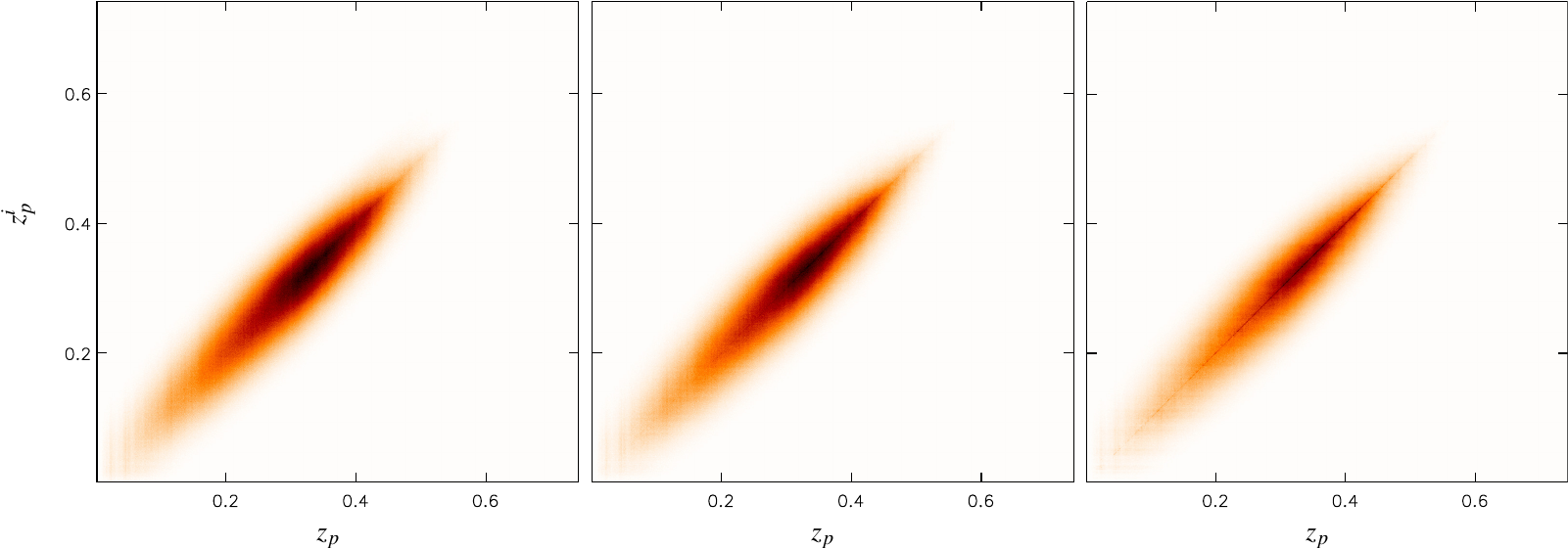}}
\caption{Correlation between the true redshift \(z_p\) and the redshifts of the \(i\)th sample \(z^i_p\) for three successive samples, tenth sample (left panel), 100th sample middle panel and 5000th sample (right panel). It can be seen that in successive samples the correlation gets stronger as the galaxies move towards their true locations.}
\label{fig:burn_in_zdisp}
\end{figure*}

The computational advantage of sampling each galaxy redshift individually can be immediately realized. Even if some individual redshift proposals are rejected, other redshift samples will be accepted leading to an overall larger acceptance probability for the entire set \(\{z_p\}\), compared to the scenario where  \(\{z_p\}\) would be sampled as a whole. Additionally, the multidimensional sampling approach can be reduced to a one dimensional sampling problem for each sampling step described in equation (\ref{eq:multiblock}), as the redshift is explored along each line of sight.

The detailed form of  the conditional redshift posterior is (Appendix \ref{Appendix_densitypost})
\begin{eqnarray}
\label{eq:cond_redshift_posterior_pap}
{\cal P}(z_n| \{\theta_p \},\{s_i\},\{z^{obs}_p\})&\propto&  \left | \left .\frac{\partial r(z)}{\partial z}\right |_{z=z_n}\, r(z_n)^2 \right | \nonumber \\
& & \times\, \sum_i W(\vec{x}_i-\vec{x}_n)\, \left(R_i \bar{N}(1+B(s)_i)\right) \nonumber \\
& & \times\, {\cal P}(z^{obs}_n|\theta_n , z_n ,\{s_i\})\, . \nonumber \\
\end{eqnarray}
where \(r(z)\) is the radial co-moving distance as a function of redshift \(z\) and \(W(\vec{x})\) is the voxel kernel function, as defined in Appendix \ref{Appendix_densitypost}. Since the Metropolis-Hastings algorithm does not require a probability distribution to be normalized in order to generate samples, we omitted all normalization constants in equation \ref{eq:cond_redshift_posterior_pap}.

The first factor in equation (\ref{eq:cond_redshift_posterior_pap}) is just the Jacobian of the coordinate transformation from spherical redshift to Cartesian co-moving coordinates, which is required since the density field is defined in co-moving Cartesian coordinates.
The second factor, is essentially the intensity function for the spatial Poisson process, as described in section \ref{log_normal_model}. It reflects the naive intuition, that the probability of finding a galaxy at a given spatial location should be related to the underlying density field. Also note, that the conditional redshift posterior distribution accounts for non-linear and non-local galaxy biasing \(B(s)_i\), as well as for the systematics of the survey under consideration via the survey response operator \(R_i\), which incorporates survey geometry and selection functions.
The last factor in (\ref{eq:cond_redshift_posterior_pap}) is the photo-z likelihood encoding the information about the redshift \(z^{obs}_n\) for the \(n\)th galaxy determined by a a photo-z estimator. Since galaxy redshifts are determined separately for individual galaxies it is reasonable to assume that these likelihoods are independent of the observed redshifts of all other galaxies, meaning the redshift likelihood for an individual galaxy does not depend on the redshift observations of other galaxies.

We  emphasize that so far we made no specific assumptions regarding the photo-z  likelihood. The aim of this work is to find a physically meaningful way of increasing the accuracy of redshift inference once all other, color or spatial information has been used. This is  achieved by the two first factors in equation (\ref{eq:cond_redshift_posterior_pap}) which could be considered as a redshift prior once the underlying density field is known. Thus, the joint sampling approach for the density field and the galaxy redshifts described in section \ref{method} provides us with samples of the joint distribution. Discarding the density fields then amounts to a Monte-Carlo marginalization over the underlying three dimensional density fields and provides us with the redshift posteriors for all galaxies conditional only on the data and the isotropy prior.

It is important to remark, that since no specific assumptions on the redshift likelihood are required, and since we are using a sampling approach, our method will in principle work with any specified redshift likelihood, or input photo-z pdf. Our method is therefore intended to  be used as a complement to any photo-z estimator which use galaxy-specific information such as a photometry or morphology to provide a photo-z pdf.

For this reason, our method is highly flexible and  therefore broadly applicable to various different scenarios and galaxy observations.   
\begin{figure*}
\centering{\includegraphics[width=0.4\textwidth,clip=true]{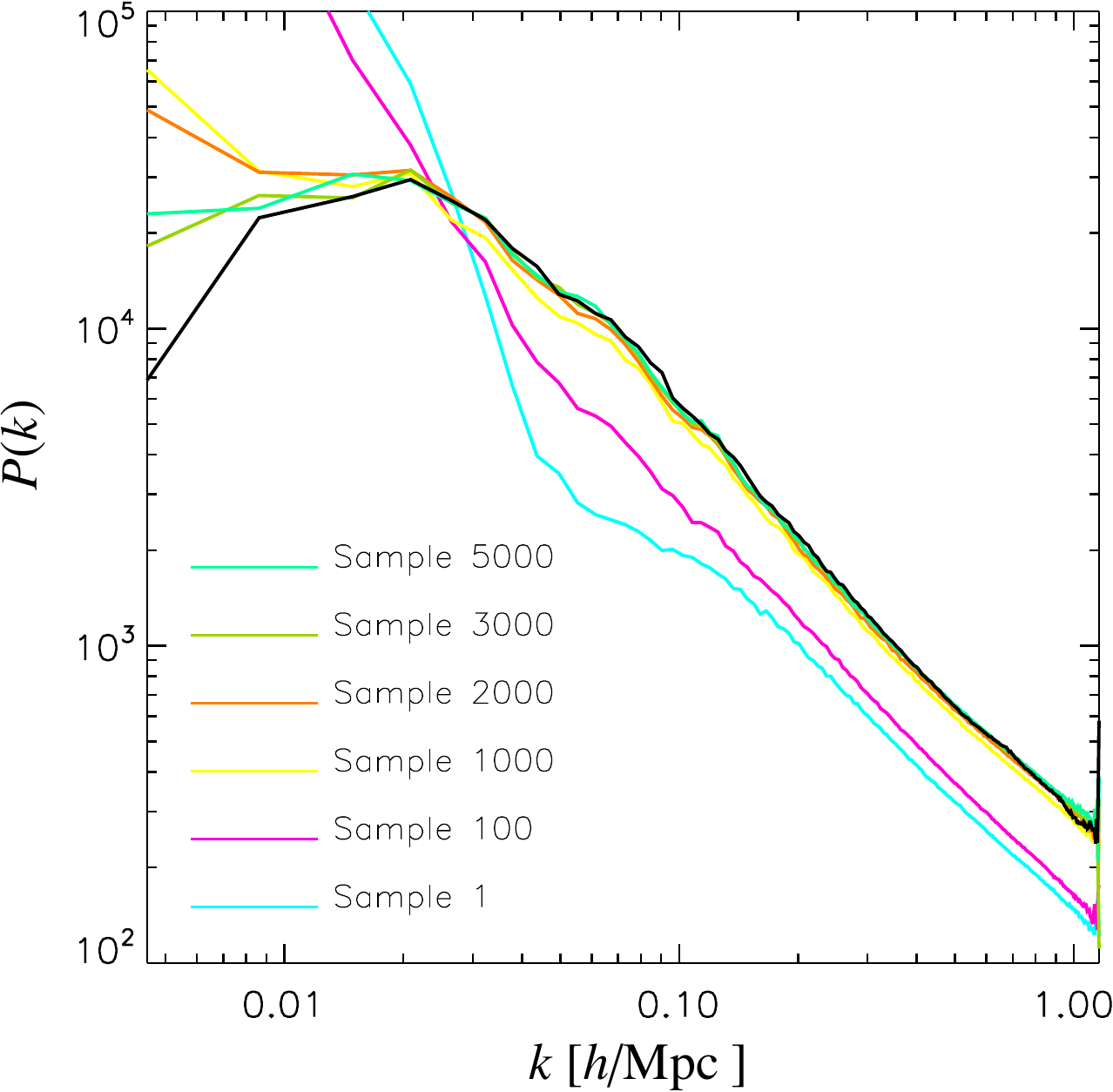}}
\caption{Power-spectra \(P(k)\) of successive samples during the burn-in phase. It can be seen that the samples move towards the true power-spectrum (black curve) of the matter field realization.}
\label{fig:burn_in_spec}
\end{figure*}

\section{Generating Mock observations}
\label{mock_observations}
In the previous sections we presented the derivation and the implementation of our method. Here we will describe the generation of artificial mock data sets, which will be used later to test our method.

\subsection{Generating galaxy redshifts}
\label{MOCK_Observation}
The generation of a artificial mock galaxy survey  will closely follow the description in \citet[][]{JASCHE2010HADESMETHOD}.
Since galaxies trace the three dimensional matter distribution and since in this work we are interested in recovering the true redshift locations of galaxies from redshift surveys with uncertain radial locations of galaxies, we prefer using a density field simulated from an N-body simulation rather than a log-normal realization. Such a density field permits us to test our method in a more realistic scenario, since the mock density field contains structures, such as voids, filaments and clusters.

For this reason, we  build our mock galaxy observation on a matter field realization calculated by the tree-PM code GADGET-2 \citep[][]{GADGET2005}, starting from Gaussian initial conditions. The details of this simulation are described in \citet[][]{CRISSIMULATION2010}. 

We estimated the simulated matter density field on a \(256^3\) Cartesian equidistant grid with side length of \(1200\) Mpc, using the Cloud in Cell (CIC) kernel and the super-sampling procedure described in \citet[][]{Jasche_2008} to account for aliasing effects. Note that we do not take peculiar velocities  into account when mapping into redshift space, since any such distortion will be negligible for photometric survey data.

Given this density field we can then generate Poisson realizations, taking into account several survey properties such as noise, survey geometry and selection effects. In figure \ref{fig:mock} we show slices through the true underlying matter field realization and the simulated redshift survey.

The survey properties are described by the galaxy selection function \(F_j\) and the observation Mask \(M_j\) where the product
\begin{equation}
\label{eq:RESPONSE_OPERATOR}
R_j=F_j\,M_j
\end{equation} 
yields the linear response operator \(R_j\) of the galaxy survey at the \(j\)th volume element.

The toy selection function, used in our tests, is given by
\begin{equation}
\label{equation:selfunc}
F_j = \left \{\begin{array}{c} (\frac{1}{\left(r_j+1-r_{th}\right)^{\beta}}) \, \, \, \,  \mbox{			for \( r_j>=r_{th} \)} \\ 1 \, \, \, \,  \mbox{			else} \end{array}\right .
\end{equation} 
where \(r_j\) is the co-moving distance from the observer to the center of the \(j\)th voxel. For our simulation we chose parameters \(\beta=0.25\) and \(r_{th}=100{\unit{Mpc}}\).
The observation Mask \(M_j\) is the same as described in \citet[][]{JASCHE2010HADESMETHOD}. It was designed to approximate the most prominent features of the Sloan Digital Sky Survey geometry \citep[][]{SDSS7}.
As a result of the Poisson sampling of the density field we obtain galaxy number counts \(N_{j}\) at each grid-point in the three dimensional domain. 

Next, in order to assign redshifts to the galaxies, all the \(N_{j}\) galaxies inside the \(j\)th volume element are uniformly distributed over the volume element to offset them from the center position. This process does not destroy the Poissonian process by which the galaxy sample was realized, since it does not change the number of galaxies inside the volume element.
Given an assumed cosmology we can then calculate the redshift \(z_p\) and the two angular positions \(\theta_n\) for each galaxy. In particular, we assumed a standard \(\Lambda\)CDM cosmology corresponding to the following set of cosmological parameters (\(\Omega_m=0.24\), \(\Omega_{\Lambda}=0.76\), \(\Omega_{b}=0.04\), \(h=0.73\), \(\sigma_8=0.74\), \(n_s=1\) ).

At the end of this procedure we obtained a set of real-space galaxy positions \(\{\theta_n, z_p\}\) for \(N_{gal}=21340276\) galaxies.

\begin{figure*}
\centering{\includegraphics[width=1.0\textwidth,clip=true]{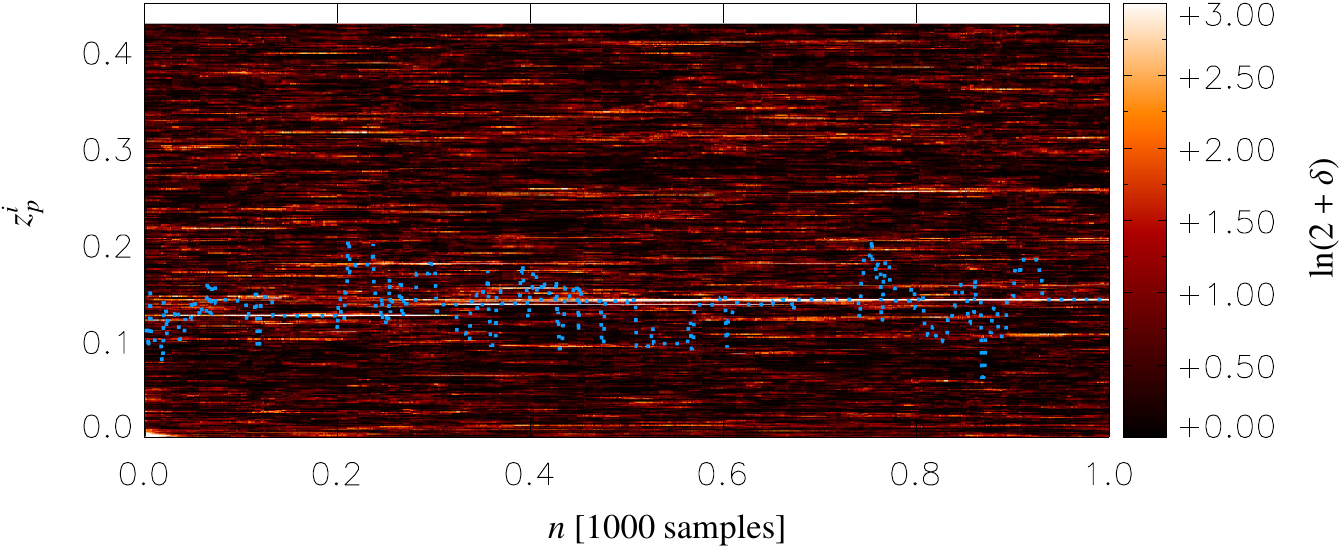}}
\caption{The plot shows the trajectory of a randomly selected galaxy along the line of sight as a function of the sampling step \(n\) in the Markov chain (blue dashed line). The background displays the density field along the line of sight in each sampling step. It can be seen that the galaxy jumps preferentially between different regions of higher density in the range allowed by the redshift likelihood from a photo-z estimator.}
\label{fig:stream_image}
\end{figure*}

\subsection{Introducing redshift uncertainties}
\label{introduce_redshift_uncertainty}
In the previous section we described how to obtain a set of unperturbed galaxy redshifts given a specific density field. Here we will describe how to generate the observed photo-z uncertainties. 

Note, that until now we have not specified a specific redshift likelihood. The method presented in this work is general, and does not need to make any specific assumptions about the functional shape of the redshift likelihood. This is due to our sampling approach which permits us to use any desired redshift likelihood.

In this work we will assume that the uncertainty in the observed redshifts is due to the observational strategy, i.e.. due to the instrument. The redshift likelihood is therefore independent of the underlying density field
\begin{eqnarray}
\label{eq:redshift_likelihood_a}
{\cal P}(z^{obs}_n|\theta_n , z_n ,\{s_i\}) &=& {\cal P}(z^{obs}_n| z_n ) \, .
\end{eqnarray}
Although it is possible to use an arbitrary redshift likelihood, here we will employ a truncated Gaussian distribution given as
\begin{eqnarray}
\label{eq:redshift_likelihood_b}
{\cal P}(z^{obs}_n|z_n) &=&\frac{\Theta\left(z^{obs}_n \right)\, {\rm{e}}^{-\frac{1}{2}\frac{\left(z^{obs}_n-z_n\right)^2}{\sigma_z^2}}}{\sqrt{\frac{\pi}{2}\sigma_z^2}\,\left({\rm{erf}}\left(\frac{z_n}{\sqrt{2\,\sigma_z^2}}\right)+1\right)}  \, ,
\end{eqnarray}
where \(\sigma_z\) is the photo-z dispersion. 
This simple distribution permits us to test our method and to estimate its general performance. Given the true redshift \(z_n\) generated via the procedure described in section \ref{MOCK_Observation} and by specifying a desired redshift dispersion \(\sigma_z\), the observed redshift \(z^{obs}_n\) can be generated by drawing random samples from the truncated Gaussian distribution, given in equation (\ref{eq:redshift_likelihood_b}), via rejection sampling.
We distort the redshift of each individual galaxy following this recipe.

For the test case considered in this work we treat all galaxies with the same \(\sigma_z= 0.03\). Since we are able to treat each galaxy individually, there is no problem in specifying a different photo-z likelihood for every galaxy. The possibility of modeling heterogenous data sets with photo-z pdfs that differ from galaxy to galaxy  is very interesting since it allows incorporating detailed uncertainty information from photo-z estimators which could take into account color or morphological information. It would also be possible to  merge data sets with different redshift accuracies, including spectroscopic surveys. We will explore these promising ideas in future work.

The process of generating a galaxy distribution with redshift distortions corresponding to the likelihood given in equation (\ref{eq:redshift_likelihood_b}) is visualized in figure \ref{fig:mock}. It can be seen that given a redshift dispersion of \(\sigma_z= 0.03\) structures along the line of sight are smeared out on a scale of roughly \(\sim 100\) Mpc.

In the following we will apply our method to this mock survey.

\section{Testing}
\label{TESTING}
In this section we will apply our method to our mock survey in order to estimate its performance. In particular, we focus on the convergence behavior of the sampler which determines the efficiency of the method in a realistic setting.

\subsection{Testing convergence and correlations}
\label{convergence}
The Metropolis Hastings Sampler is designed to have the target distribution, in our case the joint posterior distribution, as its stationary distribution \citep[see e.g.][]{metroplis,hastings,NEAL1993}.
For this reason, the sampling process will provide us with samples from the specified joint posterior distribution after a sufficiently long burn-in phase. However, the theory of Metropolis Hastings sampling by itself does not provide any burn-in criterion.
Therefore, this initial burn-in behavior has to be tested using numerical experiments.

Burn-in manifests itself as a systematic drift of the sampled parameters towards the true parameters from which the artificial data set was generated. We  monitor it by following the evolution of parameters in subsequent samples. 

As an example, in figure \ref{fig:burn_in_z} we show slices through galaxy number counts calculated from successive redshift samples. While initially the galaxy distribution exhibits prominent redshift distortions, successive samples iteratively correct the galaxy positions towards their true real-space positions. Visually, already the 100th sample does not seem to be contaminated by redshift distortions. In order to demonstrate that galaxies are really moving towards their true real-space positions during the burn-in phase, we can study the correlation of the sampled redshifts with the true galaxy redshifts for successive samples. The results of this test is presented in figure \ref{fig:burn_in_zdisp}. Here it can be seen, that the correlation between the sampled and the true redshifts tightens throughout successive samples.  
Figure \ref{fig:burn_in_zdisp} also clearly demonstrates the ability of our method significantly to enhance the constraints on the true galaxy redshifts from photo-z  surveys.

These aforementioned diagnostics particularly focus on monitoring the burn-in behavior of the redshift samples. In order to probe the evolution of density samples we monitor power-spectra estimated from successive density samples. 
As the galaxies move towards their true locations these successive power-spectra will drift towards the power-spectrum of the underlying mock matter field realization.
These successive power-spectra are depicted in figure \ref{fig:burn_in_spec}. As can be clearly seen, successive power-spectra move towards the true power-spectrum of the underlying mock matter field realization. 
In summary these tests indicate that the initial burn-in phase requires on the order of 5000 samples.

\begin{figure*}
\centering{\includegraphics[width=0.4\textwidth,clip=true]{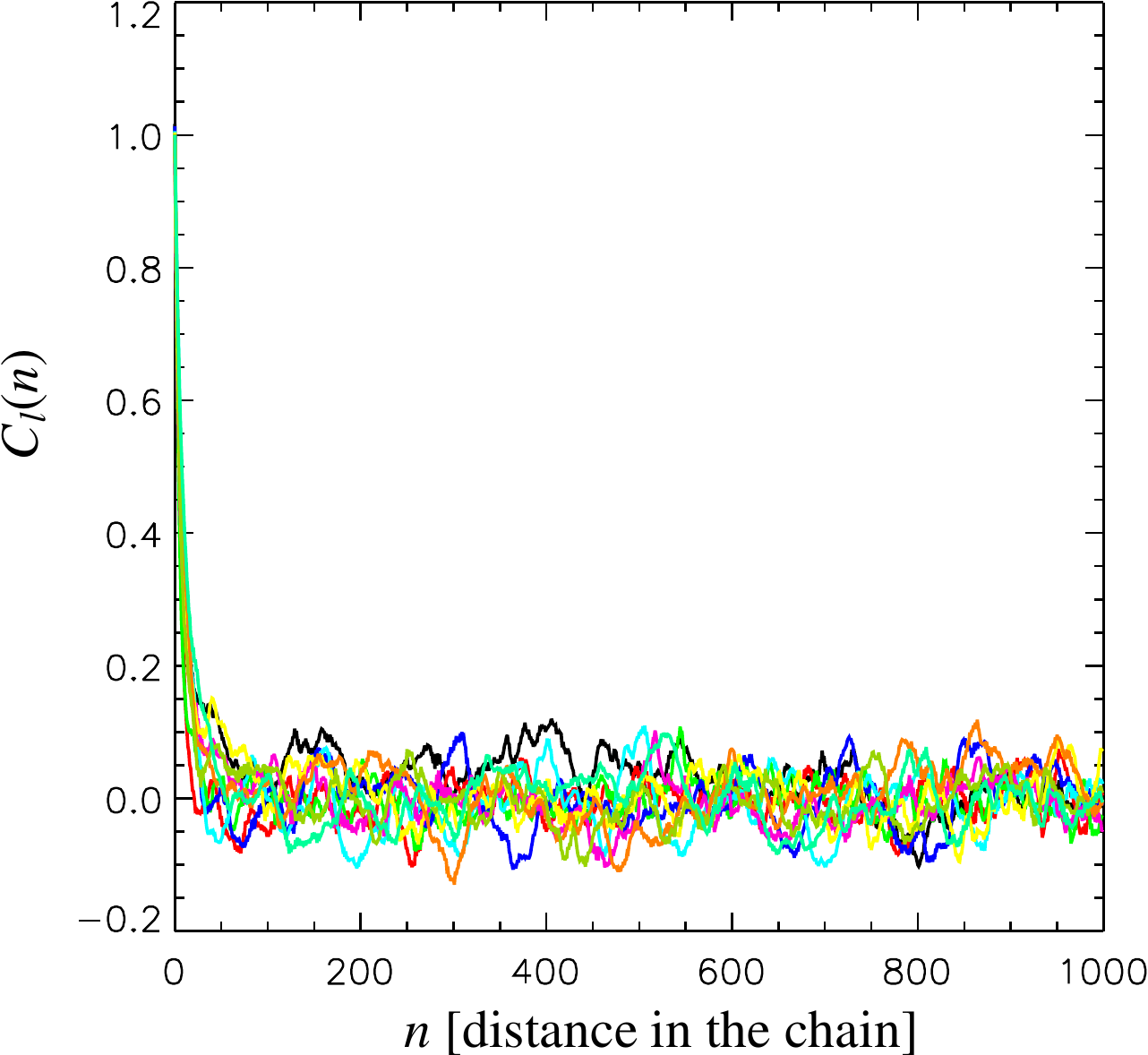}}
\caption{Correlation functions \(C_l(n)\) as a function of distance in the Markov chain \(n\) for ten randomly selected galaxies. It can be seen that the correlation length between the samples in the chain is on the order of 200 samples.}
\label{fig:corrlength}
\end{figure*}

Next we will discuss the general convergence behavior of the redshift samples in particular the correlation structure.
In general, successive samples in a Markov chain are not independent, but are correlated with previous samples. In order to estimate how many independent samples
are generated by the Markov chain one has to study this correlation effect.
This can be done by following a similar approach as described in \citep[][]{2004ApJS..155..227E} or \citep[][]{JASCHE2010PSPEC} by assuming all individual galaxy redshifts \(z^i_l\) in the Markov chain to be independent and estimating their correlation in the chain by calculation of their autocorrelation function
\begin{equation}
\label{eq:CORR_COEFF}
C_l(n) =\left \langle  \frac{z^i_l-\left \langle z_l\right \rangle}{\sqrt{\mathrm{Var}( z_l)}} \frac{z^{i+n}_l-\left \langle z_l\right \rangle}{\sqrt{\mathrm{Var}( z_l)}} \right \rangle \, ,
\end{equation}
where \(n\) is the distance in the chain measured in iterations.
In figure \ref{fig:corrlength} we show the autocorrelations of redshift samples for several randomly selected galaxies.
We can further define the correlation length of the redshift sampler as the distance in the chain \(n_C\) beyond which the correlation coefficient \(C_l(n)\) has dropped below \(0.1\). 
As can be seen in figure \ref{fig:corrlength} the correlation length is less than 200 samples, which demonstrates the high mixing rate of the sampler.

Despite the high dimensionality of the problem considered here, these tests demonstrate that the joint posterior of galaxy redshifts and three dimensional density field can be explored in a numerically feasible way by our method.  

\section{Results}
\label{results}

In the previous section we demonstrated the ability of our method to efficiently explore the joint parameter space of the three dimensional density field and the true redshifts of galaxies from redshift surveys with uncertain radial location of galaxies.

The algorithm will identify high density regions in volume elements where the number counts of galaxies is high. In this fashion, the  information of the entire galaxy sample is used to  infer the 3D density field. In turn, the  inference of individual galaxy redshifts uses this reconstructed 3D density field as described in section \ref{red_post_dist}. Therefore information propagates from the entire galaxy sample to reduce the redshift uncertainties of each individual galaxy.  

In successive Markov samples individual galaxies will preferentially move to regions of higher density within the range allowed by the photo-z inference. For this reason, a worry might be that the galaxy get stuck and not move any further once it found a density peak.
We find that this does not occur since our burn-in and autocorrelation tests show that galaxies quickly find and explore their redshift range.

It is illustrative to follow the trajectory of an individual galaxy along its line of sight throughout successive samples. 
Generally, in each sample step both, the density field along the line of sight and the galaxy redshift, will change. In figure \ref{fig:stream_image} we show the trajectory of a random galaxy and the evolution of the corresponding density field along the line of sight for 1000 samples. It can be seen that the galaxy preferentially moves to regions with densities higher than the mean density and avoids underdense regions. Also the galaxy stays longer in regions of higher density, which correspond to regions of higher probability, than in other regions. 

This is the anticipated behavior of the Markov sampler, which is supposed to provide more samples of high probability regions and less samples in regions of lower probability in order to construct a probability density.
 
Next, we want to discuss the final scientific results of the sampling procedure. In particular, we are interested in the accuracy of the inferred galaxy redshifts.
Intuitively, the accuracy of the inferred redshifts should  depend on the environment at the galaxies' original location. Galaxies born in high density regions will  be attracted by a high density peak along their line of sight, while galaxies born in low density regions have fewer constraints. In the limit of very low density the galaxy will explore the full range allowed by the input photo-z likelihood along the line of sight. 

High density regions are better sampled by galaxies, which further decreases the uncertainty for the inferred density field at these positions, which results in an overall higher joint probability for the density field and galaxy redshifts at these locations. For this reason, redshifts of galaxies located in high density regions such as clusters or super clusters will generally be recovered more precisely than those of galaxies in regions with lower density. 

\subsection{Redshift samples}
\label{sampled_redshift_dist}
\begin{figure*}
\centering{\includegraphics[width=0.4\textwidth,clip=true]{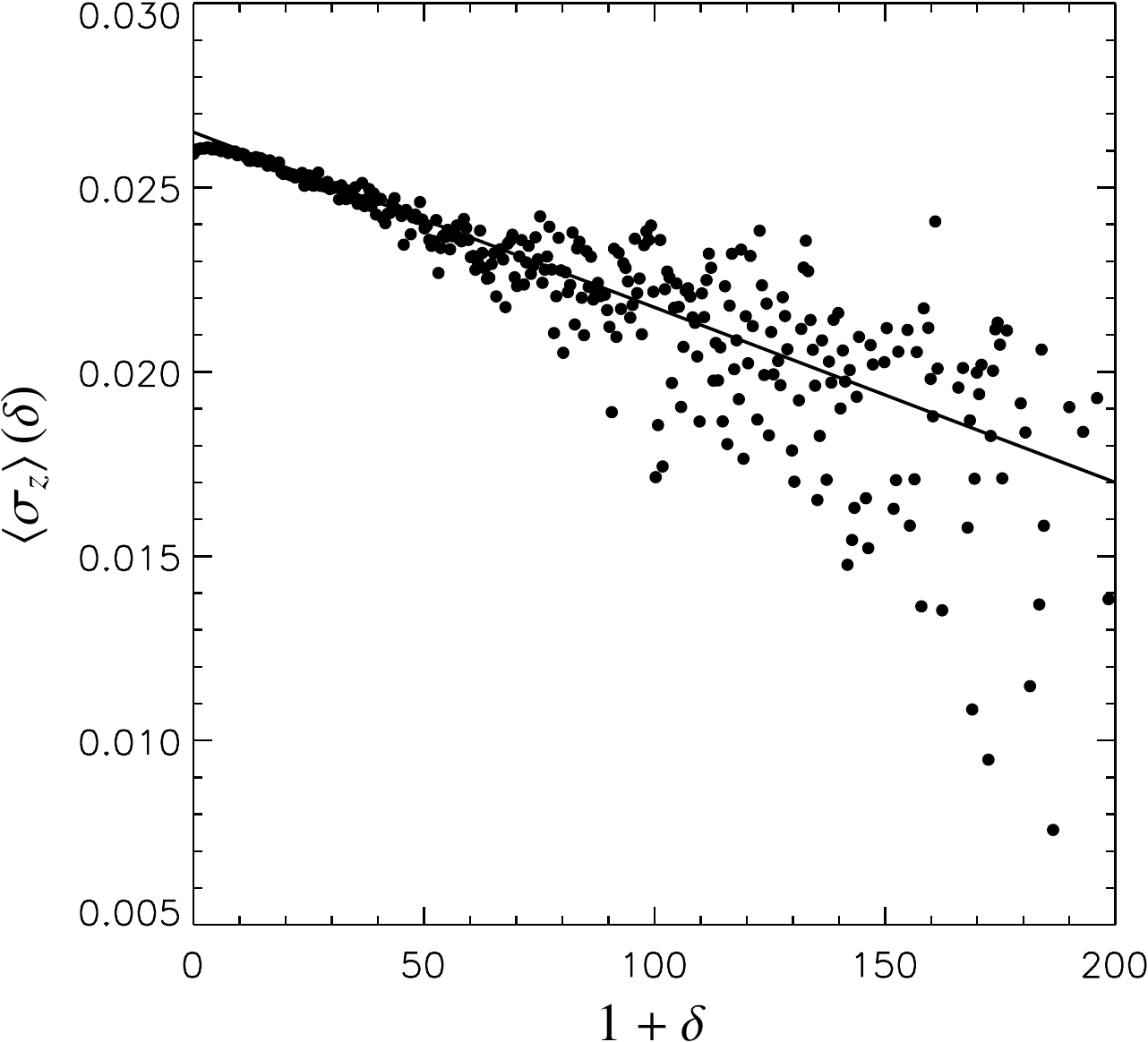}}
\caption{Mean standard deviation as a function of density contrast \(\delta\).}
\label{fig:zdensvar}
\end{figure*}

To demonstrate this effect and to see which regions can be best recovered from a redshift survey with uncertain radial locations of galaxies, we study the average redshift standard deviation as a function of the true underlying density field \(\delta_{true}\)
\begin{equation}
\label{meansigmaz} 
\langle \sigma_z \rangle(\delta_{true})=\frac{\sum_p\,\sigma_p\, K(\delta_{true}^p - \delta_{true}) }{\sum_p\,K(\delta_{true}^p - \delta_{true})} \, ,
\end{equation} 
with \(\delta_{true}^p\) being the true density contrast at true location of the \(p\)th galaxy, \(\sigma_p\) is the galaxy redshift standard deviation calculated from the samples and \(K(x)\) is the binning kernel as defined in equation \ref{equation:kernel_function}. Equation (\ref{meansigmaz}) yields the redshift standard deviation measured from the Markov chain averaged over density bins of the true underlying mock matter-field realization. This quantity is a measure of how well the true redshift can be recovered from an uncertain galaxy redshift survey.

Figure \ref{fig:zdensvar} clearly shows a strong dependence of the average redshift standard deviation on the amplitude of the true density field. Very dense structures can be recovered very well. As expected, figure \ref{fig:zdensvar} demonstrates that high density regions such as clusters can be much better recovered from uncertain redshift surveys than regions of lower density such as filaments or voids. However, on average the redshift precision has improved for all galaxies compared to the initial uncertainty.

\begin{figure*}
\centering{\includegraphics[width=0.6\textwidth,clip=true]{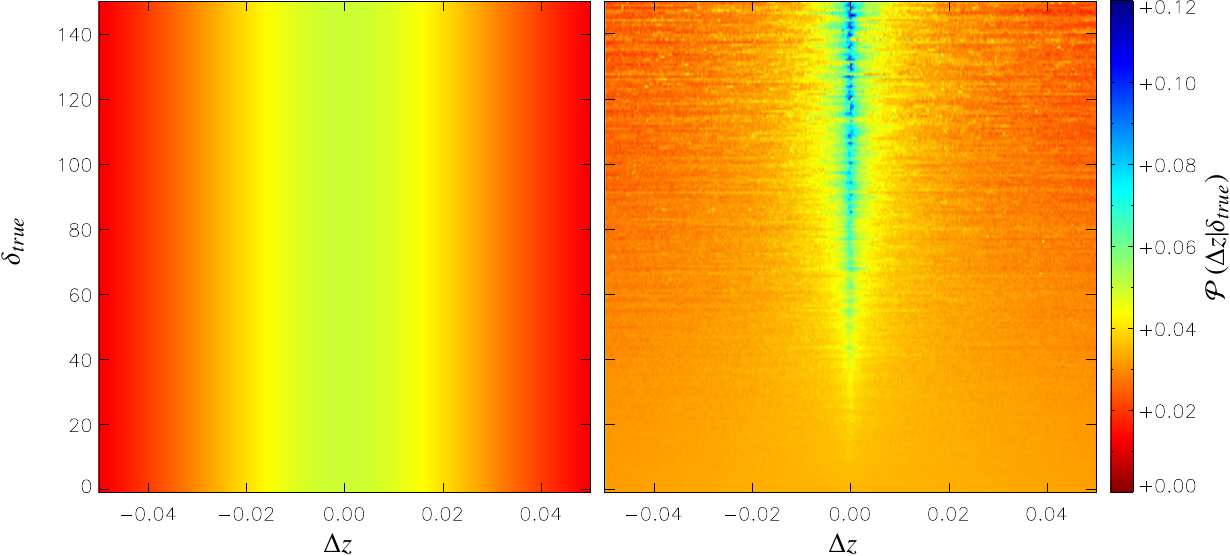}}
\caption{Probability distribution for the difference \(\Delta z\) between measured and true redshift conditional on the true underlying density field \(\delta_{true}\). The left panel shows the initial Gaussian redshift likelihood which is conditionally independent of the true underlying density field. The right panel shows the a posteriori conditional probability density distribution for \(\Delta z\) estimated from the Markov samples. It can be seen that the accuracy of inferred redshifts depends on the true underlying density field. Also note that the resulting probability distribution is non-Gaussian. }
\label{fig:PDF_DELTAZ}
\end{figure*}

The final redshift posterior distributions are generally highly non-Gaussian.
We therefore estimate the conditional probability distribution \(\mathcal{P}(\Delta z|\delta_{true})\) of redshift departures from the true redshift \(\Delta z= z_{true} - z\) conditional on the true underlying density field \(\delta_{true}\)
\begin{equation}
\label{eq_diff_redshift}
\mathcal{P}(\Delta z|\delta_{true})= \frac{ \sum^{N_{samp}}_{l=1} \sum_p \delta^D(\delta_{true}^p - \delta_{true}) \, \delta^D(\Delta z - (z_{true}^p-z_l^p))}{\sum^{N_{samp}}_{l=1} \sum_p \delta^D(\delta_{true}^p - \delta_{true})}\, ,
\end{equation}
which is the probability of finding the departure \(\Delta z\) in a region with density contrast \(\delta_{true}\) after our method has been applied. Initially this distribution coincides with the redshift likelihood, which is independent of the density field and is basically a Gaussian distribution with redshift dispersion \(\sigma_z=0.03\). The comparison between the initial distribution \(\mathcal{P}(\Delta z|\delta_{true})\) and that obtained by our method is shown in figure \ref{fig:PDF_DELTAZ}. This plot demonstrates that although the initial redshift likelihood is independent of the true underlying density field, the final redshift posterior distribution strongly depends on it. 

Beyond measures of accuracy, it is instructive to inspect the redshift posteriors we derive for individual galaxies. 
In figure \ref{fig:PDF_Z} we show some randomly selected redshift posterior distributions. In general the inferred redshift posterior distributions are  non-Gaussian and multimodal. Each mode corresponds to a possible association of the galaxy with peaks in the density distribution. The area under each mode measures the probability for the galaxy to be associated with that density peak. In several cases these modes sit on top of a broader distribution which  for the residual probability that the galaxy is not associated with any of these peaks. It is clear that by any reasonable measure of information, such as Shannon's negentropy, these pdfs are far more informative regarding the redshift of the galaxy than the original photo-z likelihoods.

\subsection{The density field}
\label{densityfield}
In this section we will discuss the inferred three dimensional density field. As already described in the previous section, the accuracy of the inferred redshifts depends on the amplitude of the true underlying density field. In particular, the higher the underlying density the more accurate we can identify the true location of galaxies. For this reason, we expect a similar behavior for the recovered three dimensional density field. For example, we expect to recover high density regions more accurately than low density regions. 
However, the corresponding joint uncertainty of redshift and density estimates can be accurately estimated from the Markov samples. As an example we calculated the ensemble mean and variance for the density field. The results are presented in figure \ref{fig:mean_variance_mask} where we show three slices through the ensemble mean density and the ensemble variance field from different sides. The results demonstrate that high density regions are much better localized than lower density or even under dense regions. In particular, the ensemble mean density field shows elongated structures along the line of sight for under dense objects such as voids. This result was anticipated. It reflects the fact that these objects are poorly constrained by the data.

To further quantify this  visual impression and to estimate the accuracy of the inferred density field we study point to point comparisons between the true underlying density field and the recovered ensemble mean density field smoothed on different scales. The density fields are smoothed with a spherical top hat filter in Fourier space with different filter radii \(k_{th}\). We can then calculate the corresponding correlation coefficient for the different filter thresholds \(k_{th}\)
\begin{equation}
r(k_{th})=\frac{\left\langle \delta_{true}^{k_{th}} \, \langle\delta \rangle^{k_{th}} \right \rangle}{ \sqrt{\langle \left(\delta_{true}^{k_{th}}\right)^2 \rangle}\,\sqrt{\langle \left(\langle\delta \rangle^{k_{th}}\right)^2 \rangle}}\, ,
\end{equation}
where the superscript denotes that the field has been filtered with a top hat filter corresponding to \(k_{th}\). The result of this calculation is presented in figure \ref{fig:dens_corr_coeff}.
To further study the dependence of the accuracy of the inferred density field on the true underlying density field, we will also calculate the correlation coefficients only  between voxels in the true and the inferred ensemble mean density for which \(\delta_{true}> \delta_{th}\). We can therefore study the correlation between true and recovered density field as a function of smoothing scale and density amplitude.
As demonstrated by figure \ref{fig:dens_corr_coeff}, the accuracy for the inferred structures in the density field strongly depends on the amplitude of the true underlying density field. As expected, high density objects can generally be much better constrained than low density or even under dense regions. It is interesting to point out, that high density objects still show a reasonably high correlation even at the shortest scales resolved by the grid (\(\sim 4.69\) Mpc) although the data corresponds to a density field smeared out on a scale of about \(100\) Mpc due to a redshift variance of \(\sigma_z=0.03\).
These results are therefore in agreement with the results for the estimated redshifts presented in the previous section. In the absence of additional information or data the inference of redshifts and three dimensional density field strongly depends on the amplitude of the true underlying density field.

\section{Discussion and conclusion}
\label{discussion}
Present and upcoming photometric galaxy surveys  provide us with (tens of) millions of galaxies  probing the density field out to high redshifts.  These enormous numbers of galaxies can only be obtained at the expense of low accuracy in the measurement of the radial locations of galaxies along the line of sight. It has been demonstrated that these redshift uncertainties will generally influence the inference of cosmological signals such as the baryon acoustic oscillations, cluster counts or weak lensing.
For this reason, accurate treatment of redshift uncertainties is required in order to use these surveys to full effect.

In this work, we presented a new, fully Bayesian approach which improves the accuracy of inferred redshifts  \textit{even if all the available information based on photometry and morphology of individual galaxies has been used to estimate the photo-z}. Specifically our method performs a joint inference of the three dimensional density field and the true radial locations of galaxies from surveys with highly uncertain redshift measurements. This joint approach permits us to impose physically motivated constraints on the redshift and density inference, which increase the accuracy of the inferred locations of galaxies, essentially super-resolving high-density regions in redshift space.

Where does the additional information come frome? The isotropy and 2-D correlation prior allow improving the inference of three-dimensional structure from the highly precise angular information in the radial direction. If the projected density of tracers on the sky shows a peak then it is likely that this corresponds to one or more three-dimensional peaks along the line of sight. Many galaxies with imprecise redshifts populate the line of sight. Galaxies in high density regions can be associated with structures that are consistent with their photo-z likelihoods and the observed two-dimensional density. Each such structure has a relatively accurate redshift derived from the averaged redshift of the galaxies it contains. 

In this work we employed a simplified picture of galaxy formation. The Poisson model of galaxy formation serves to connect  the isotropy prior on the underlying density field and the three dimensional locations of galaxies. As demonstrated in section \ref{red_post_dist}, in this picture, the probability of finding a galaxy at a given radial position is proportional to the intensity of the spatial Poisson process. As can be seen from equation (\ref{eq:cond_redshift_posterior_pap}), ignoring survey systematics and galaxy biasing, the probability of finding a galaxy at a given location is essentially proportional to the density, as would be expected intuitively. Further, equation (\ref{eq:cond_redshift_posterior_pap}) also demonstrates that our method can accurately account for survey systematics and can in principle be combined with a non-linear and non-local galaxy biasing model.
\begin{figure*}
\centering{\includegraphics[width=0.8\textwidth,clip=true]{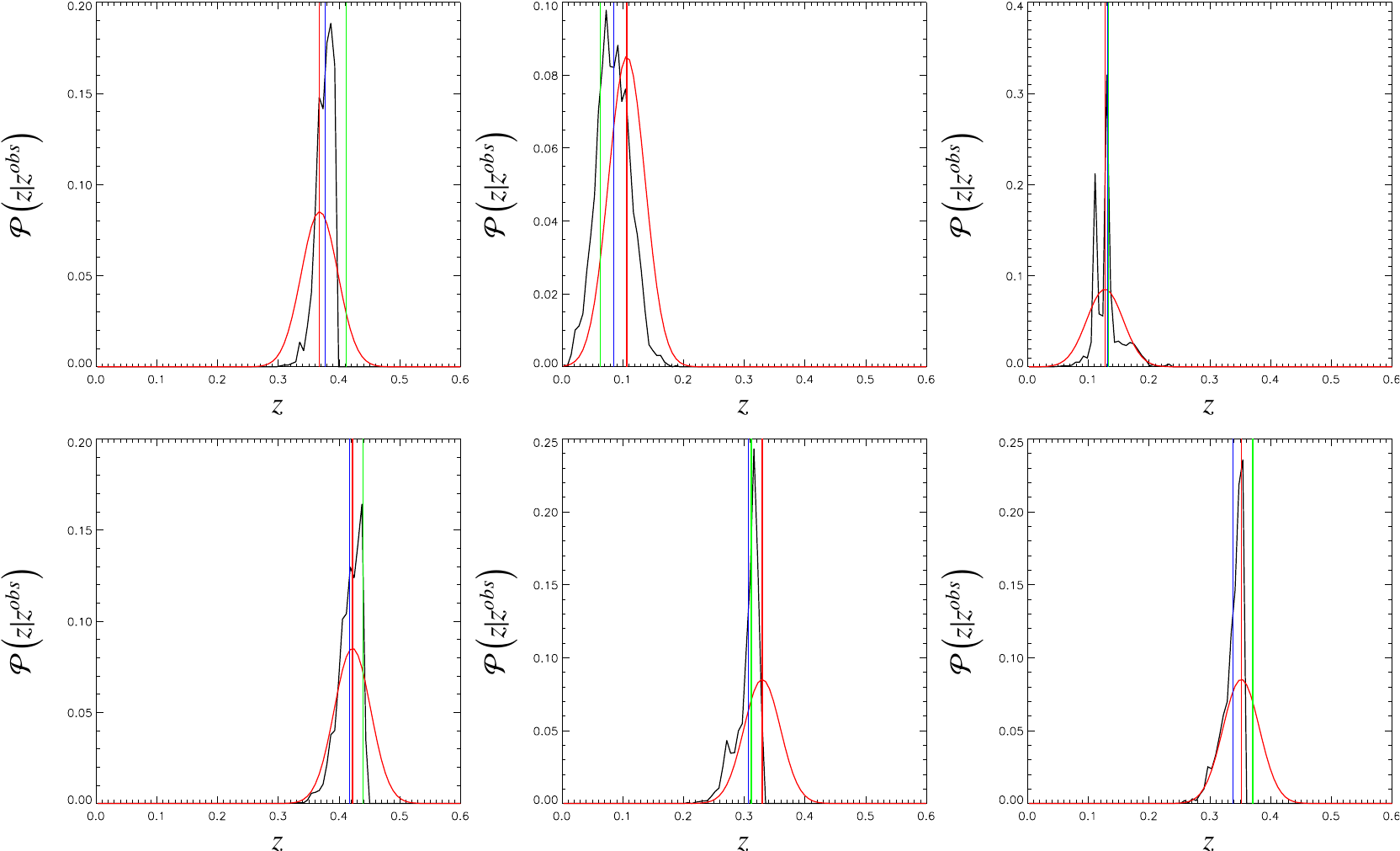}}
\caption{Randomly selected marginalized redshift posterior distributions estimated from the Markov samples (black curves). The red vertical lines denote the true redshift of this specific galaxy and the red curve shows the distribution from which the observed redshift was drawn. The green vertical lines are the observed redshifts and the blue vertical lines denote the ensemble mean redshift estimated from the samples.}
\label{fig:PDF_Z}
\end{figure*}

The joint inference of radial galaxy positions and the three dimensional density field as well as the quantification of joint uncertainties require exploring the posterior density  conditional on observations. Doing so via Markov Chain Monte Carlo methods is a numerically challenging problem due to the high dimensionality, non-Gaussianity and non-linearity of the problem.

We show that the complex sampling procedure can be split up into many simpler sub-sampling steps within the framework of a multiple block Metropolis-Hastings algorithm. In particular this permits us to split the problem into the two tasks of individually sampling three dimensional density fields, conditional on the true galaxy locations, and sampling galaxy redshifts, conditional on the density field. This sampling approach is depicted in figure \ref{fig:flowchart}.
We explore  the log-normal Poissonian posterior distribution for the three dimensional density field based on the galaxy distribution using the HADES algorithm\citet{JASCHE2010HADESMETHOD}. In this work we add sampling the redshift posterior distribution as shown as the first step in equation (\ref{eq:MARKOV_STEPS}).

In the Poissonian picture of galaxy formation the locations of individual galaxies are conditionally independent of the locations of all other galaxies once the underlying density field is given. We demonstrate this intuitively plausible fact in section \ref{red_post_dist}. This fact greatly simplifes the numerical implementation of the redshift sampler, since it allows treating each galaxy individually leading to a numerically efficient,  parallel sampling procedure despite the high dimensionality and non-linearity of the problem.

We extensively tested our method with artificial mock galaxy observations to evaluate the overall performance of the algorithm in a realistic scenario with tens of millions of galaxies. The generation of the artificial galaxy survey is described in section \ref{mock_observations}. In this test all galaxy positions have been distorted corresponding to a truncated Gaussian with equal redshift dispersion for all galaxies. In particular, we used a redshift dispersion \(\sigma_z=0.03\) which amounts to an uncertainty along the line of sight of about \(\sim 100\) Mpc. However, it should be noted that since our method treats each galaxy individually, it is generally possible to take into account the photo-z information and uncertainties appropriate for each individual galaxy.

The primary goal of the tests we conducted was to establish the numerical feasibility of our method. In particular we focused on evaluating the burn-in and  convergence behavior of the multiple block Metropolis-Hastings sampler. We studied the burn-in behavior of the sampler by monitoring the systematic drift of various inferred quantities towards their true values throughout subsequent sampling steps. As described in section \ref{convergence}, following the evolution of inferred power spectra, shown in figure \ref{fig:burn_in_spec}, and the flow of galaxies towards their true locations, as demonstrated by figure \ref{fig:burn_in_zdisp}, indicates a burn-in time of about 5000 samples.
Further, we studied the correlation length of individual galaxy redshifts across the Markov Chain. In particular figure \ref{fig:corrlength} demonstrates that the correlation length is typically less than 200 samples. These tests clearly demonstrate the feasibility of the method proposed in this work. 

As a scientific result the method provides improved photo-zs for each galaxy and the three dimensional density field as well as the corresponding joint uncertainties.
In section \ref{sampled_redshift_dist} we discuss the results for the inferred galaxy locations. As was anticipated, the accuracy of inferred redshifts depends on the density amplitude of the true underlying density field. Figure \ref{fig:zdensvar} demonstrates, that on average all galaxy redshifts have been improved with respect to the original redshift dispersion, up to a factor of 10 in high density regions.

It can clearly be seen that the accuracy of inferred redshifts for galaxies living in high density regions is generally larger than for galaxies in regions of lower density. This result is further clarified by figure \ref{fig:PDF_DELTAZ} which shows the probability distribution of departures from the true redshifts conditional on the underlying density field. Redshifts of galaxies in high density regions can be particularly well recovered from a survey with uncertain radial galaxy positions. These results correspond to our expectation, and simply reflect the fact that there are usually many galaxies in high density objects. The joint information of these many observed redshifts permits to accurately infer the position of the high density object and hence the locations of galaxies living in this object. The lower the density the smaller is the number of galaxies contributing to  a feature and hence the number of observed redshifts. As a result it is generally more difficult to constrain the locations of low density features than the location of high density features. 
\begin{figure*}
\centering{\includegraphics{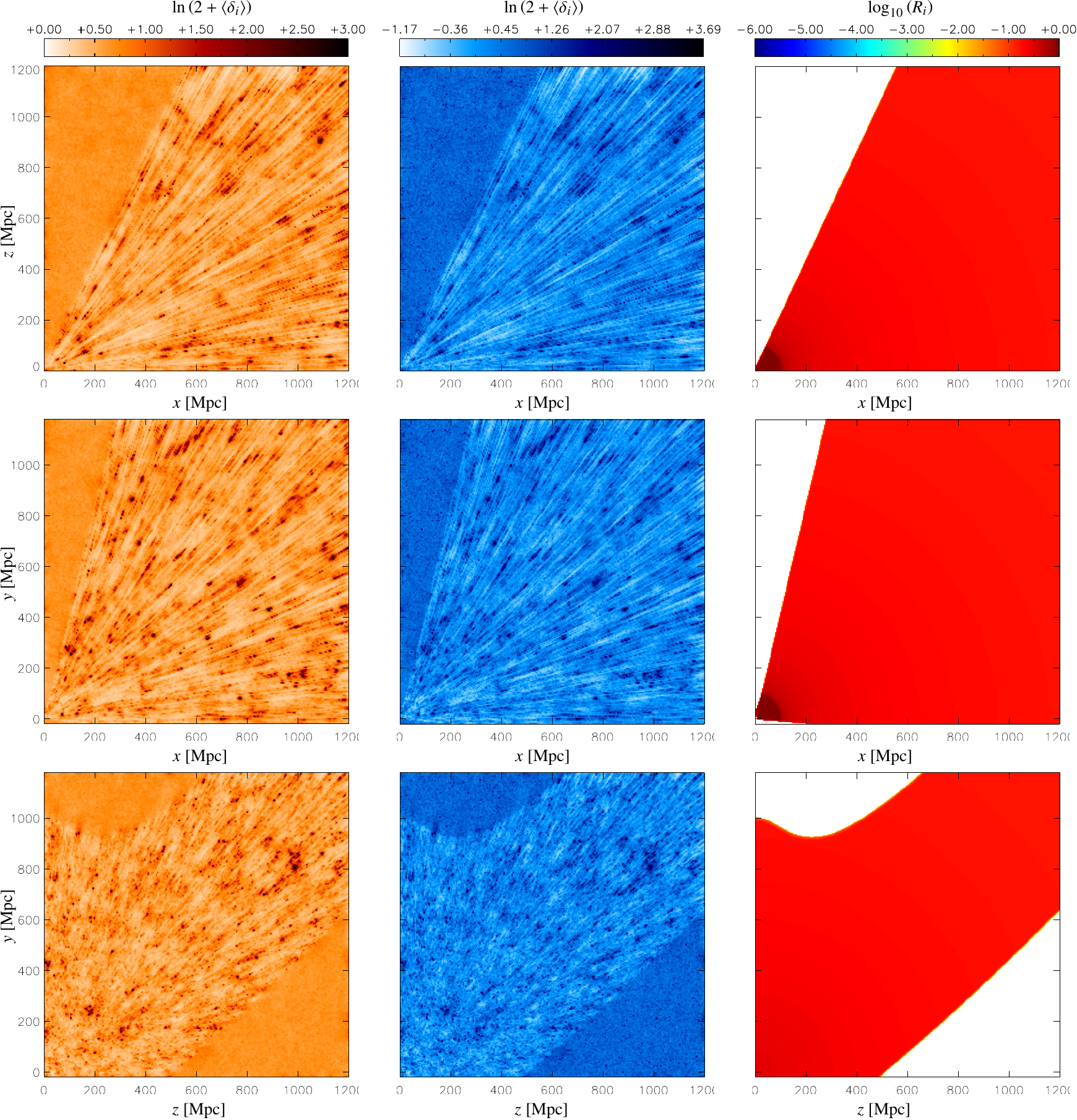}}
\caption{Three different slices from different sides through ensemble mean density (left panels), ensemble variance (middle panels) and the three dimensional response operator \(R_i\) (right panels). Especially the variance plots demonstrate, that the method accounted for the full Poisonian noise structure introduced by the galaxy sample. One can also see the correlation between high density regions and high variance regions, as expected for Poissonian noise.}
\label{fig:mean_variance_mask}
\end{figure*}

\begin{figure*}
\centering{\includegraphics[width=0.4\textwidth,clip=true]{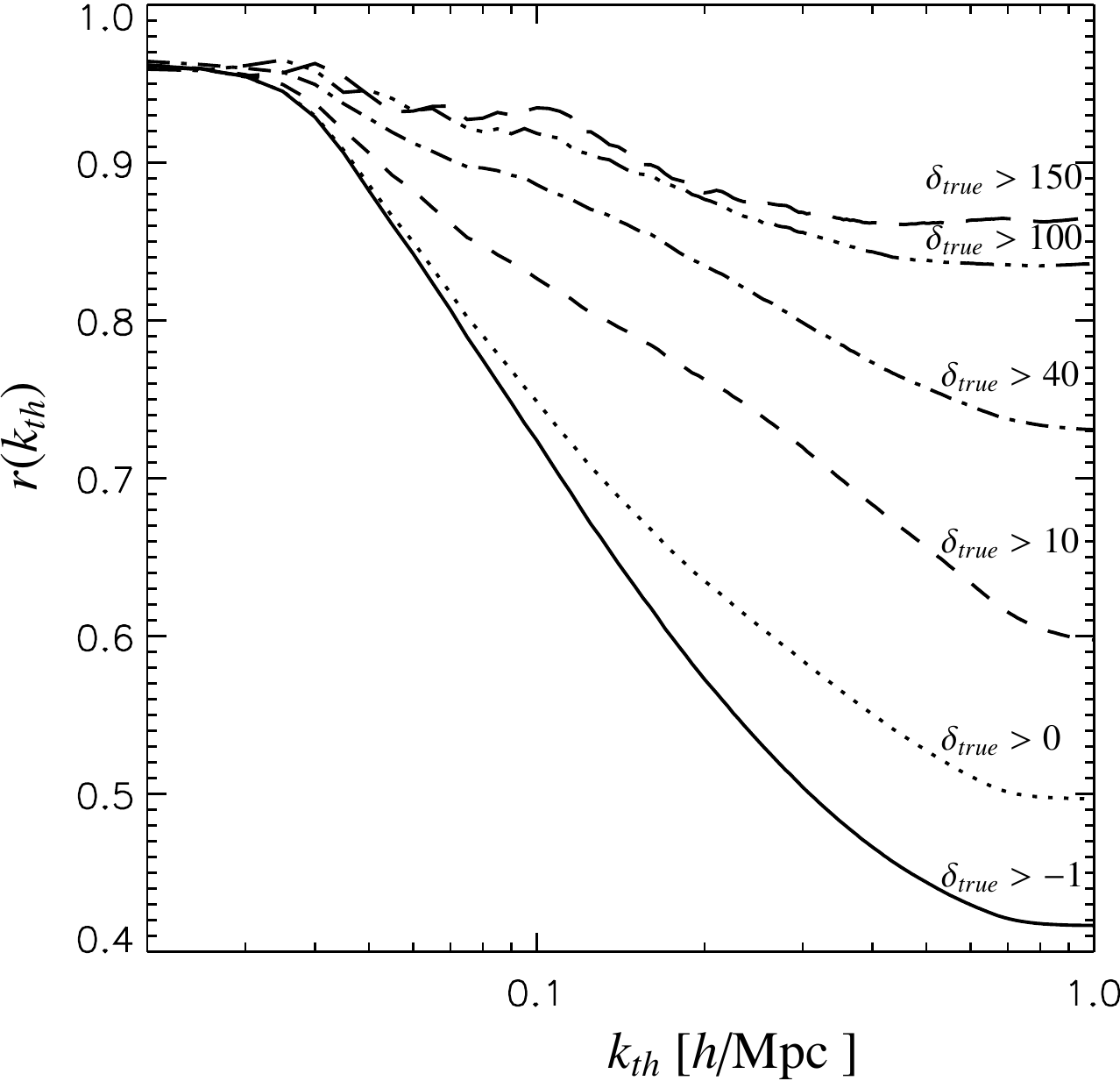}}
\caption{Correlation coefficient \(r(k_{th})\) as a function of filter scale \(k_{th}\) for different density amplitude thresholds \(\delta_{th}\) as indicated in the plot. It can be seen that the correlation between amplitudes in the true underlying density field and the inferred ensemble mean density field increases on all filter scales with increasing density thresholds \(\delta_{true} > \delta_{th}\).}
\label{fig:dens_corr_coeff}
\end{figure*}

This effect is also reflected in the inferred density field, see section \ref{densityfield}. Figure \ref{fig:mean_variance_mask} shows the ensemble mean density field and the corresponding ensemble variance estimated from the set of Markov samples. It is visually clear that high density regions can be much more accurately located than low density or even underdense regions. To quantify this impression, we studied the point to point correlation of the estimated ensemble mean with the true underlying density field for different smoothing scales and different density thresholds. The results shown in figure \ref{fig:dens_corr_coeff} clearly demonstrate that high density regions can be accurately recovered for all smoothing scales, while the correlation between ensemble mean and true density field worsens with decreasing density.
For this reason, the results of inferred density fields and galaxy redshifts are consistent and strongly depend on the density amplitudes of the true underlying density field.

We emphasize that the scientific output of our method is not a single estimate but a sampled representation of the joint posterior of three dimensional density field and galaxy positions. Any desired statistical summary as well as corresponding uncertainties can be calculated from this set of  samples. 

For example, by marginalizing over the density field we are able to provide the marginalized redshift posterior distributions, as depicted in figure \ref{fig:PDF_Z}, which permit easy propagation of  uncertainties through subsequent analysis stages.

Our method does not rely on specific assumptions on the provenance of the redshift likelihood. It is therefore applicable generally to surveys of isotropic distributions of tracers with uncertain radial positions.

In summary, the presented method is a flexible and numerically efficient addition to the analysis toolbox for large scale galaxy surveys. It improves accuracy of galaxy redshifts and three dimensional density fields from photometric galaxy redshift surveys and therefore has the potential to add substantial value to their scientific output.

\section*{Acknowledgments}
We thank Francisco S. Kitaura, Torsten A. En\ss lin and Simon D. White for useful discussions and encouraging us to pursue the project described in this work.
Further, we thank Cristiano Porciani for providing us with the simulated density field and Nina Roth for providing us with required reading routines and information on how to handle the simulation data.
Particular thanks go to Rainer Moll and Bj\"{o}rn M. Sch\"{a}fer for useful discussions and support with many valuable numerical gadgets. 
This work has been supported by the Deutsche Forschungsgemeinschaft within the Priority Programme 1177 under the project PO 1454/1-1 and by NSF grant AST 07-08849.

\bibliography{paper}
\bibliographystyle{mn2e}

\appendix

\section{Methods for obtaining redshifts from photometry}
\label{photozmethods}

The approach we present in this paper uses the output of  photo-z estimators as its input. We give a short summary of these methods and references to relevant work below. 

A variety of different methods to infer accurate redshifts from photometric galaxy observations have been proposed. These techniques can be roughly categorized as kernel regression methods, artificial neural networks and \(\chi^2\) model testing  \citet{WOLF2009}.
The class of kernel regression methods utilize sets of model realizations or spectroscopic galaxy samples to approximate the conditional probability density of a galaxy redshift conditional on photometric properties via specified kernel functions. Given this approximate probability density the non-linear mapping between a galaxies photometric properties and its redshift can be established via a kernel regression which provides the conditional expectation value of a galaxy redshift \citep[][]{NADARAYA1964,WATSON1964,BORIS2007,WANG2007,WOLF2009}.      

Artificial neural networks aim at fitting the non-linear relation between observed photometric parameters and the redshift of galaxies \citep[][]{COLLISTER2004,FIRTH2003}. In this approach, the non-linear relation between photometry and galaxy redshifts is represented by a multilayer perceptron neural network, trained by a sufficiently large training set for which both photometry and accurate redshifts are known \citep[][]{COLLISTER2004}. Generally artificial neural networks provide a single unique parameter estimate. However, if the solution is degenerate, they tend to return the most likely one \citep{WOLF2009}. It is also possible to estimate uncertainties of the inferred quantities by re sampling the input parameters from their error distributions. The outputs of a artificial neural network will then provide a probability distribution characterizing the uncertainty in the inferred quantity \citep{WOLF2009,COLLISTER2004}.   

Additionally, Bayesian redshift inference methods have been proposed in literature \citep[][]{WOLF2009,BENITEZ2000}. These models rely on \(\chi^2\) model testing, assuming a parameterized model between photometric properties and redshifts of galaxies \citep[][]{WOLF2009}. Uncertainty information on the observations then permits to determine a probability density distribution for the inferred quantity and to provide expectation values and parameter uncertainties \citep[][]{WOLF2009}. 

Improvements in photo-z determinations help to improve the technique we describe here. We emphasize that our work begins where photo-z estimation leaves off.

\section{density posterior}
\label{Appendix_densitypost}
The density posterior distribution given in equation (\ref{eq:MARKOV_STEPS}) can be written as
\begin{equation}
\label{eq:dens_post1}
{\cal P}(\{s_i\}|\{z_p\},\{z^{obs}_n\},\{\theta_n\}) = {\cal P}(\{s_i\}|\{z_p\},\{\theta_n\}) \, ,
\end{equation}
by noting that the posterior distribution is conditionally independent of the observed redshifts \(\{z^{obs}_n\}\) once the true redshifts \(\{z_p\}\) are known.
In order to relate this distribution to a log normal Poissonian distribution we would like to express it in terms of galaxy counts rather than in terms of galaxy redshifts and positions in the sky. We can therefore write equation (\ref{eq:dens_post1}) as the marginalization over all possible galaxy number count configurations \(\{N_i\}\)
\begin{eqnarray}
\label{eq:dens_post2}
{\cal P}(\{s_i\}|\{z_p\},\{\theta_n\}) &=& \sum_{\{N_i\}} {\cal P}(\{s_i\},\{N_i\}|\{z_p\},\{\theta_n\}) \nonumber \\
&=& \sum_{\{N_i\}} {\cal P}(\{s_i\}|\{N_i\})\,{\cal P}(\{N_i\}|\{z_p\},\{\theta_n\}) \, ,  \nonumber \\
\end{eqnarray}
where we assumed the conditional independence \({\cal P}(\{s_i\}|\{N_i\},\{z_p\},\{\theta_n\})={\cal P}(\{s_i\}|\{N_i\})\). Further, the number counts are defined in co-moving Cartesian coordinates, we will therefore introduce the set of corresponding co-moving Cartesian galaxy positions \(\{\vec{x}_p\}\) and marginalize over it
\begin{eqnarray}
\label{eq:dens_post3}
{\cal P}(\{s_i\}|\{z_p\},\{\theta_n\}) &=& \sum_{\{N_i\}} {\cal P}(\{s_i\}|\{N_i\})\,\int \mathrm{d}\{\vec{x}_p\} \, {\cal P}(\{N_i\},\{\vec{x}_p\}|\{z_p\},\{\theta_n\}) \, ,  \nonumber \\
&=& \sum_{\{N_i\}} {\cal P}(\{s_i\}|\{N_i\})\,\int \mathrm{d}\{\vec{x}_p\} \, {\cal P}(\{N_i\}|\{\vec{x}_p\})\nonumber \\
& & \times \,  {\cal P}(\{\vec{x}_p\}|\{z_p\},\{\theta_n\})  \, .  \nonumber \\
\end{eqnarray}
Since there exists a unique relation between the co-moving Cartesian and the true redshift coordinates we can write
\begin{eqnarray}
\label{eq:cart_coordinates}
{\cal P}(\{\vec{x}_p\}|\{z_p\},\{\theta_n\}) &=& \prod_p \delta^D \left(x^1_p - r(z_p)\,\mathrm{cos}(\delta_p)\, \mathrm{cos}(\alpha_p)\right)\nonumber \\
& & \times \, \delta^D \left(x^2_p-r(z_p)\,\mathrm{cos}(\delta_p)\, \mathrm{sin}(\alpha_p) \right)\nonumber \\
& & \times \, \delta^D \left(x^3_p-r(z_p)\,\mathrm{sin}(\delta_p) \right) \, ,
\end{eqnarray}
with \(\alpha_p\) and \(\delta_p\) being the right ascension and declination of the \(p\)th galaxy respectively.
By introducing the vectors \(\vec{y}_p\) as
\begin{eqnarray}
\label{eq:cart_coordinates}
\vec{y}_p= r(z_p) \left [\begin{array}{c} \mathrm{cos}(\delta_p)\, \mathrm{cos}(\alpha_p) \\  \mathrm{cos}(\delta_p)\, \mathrm{sin}(\alpha_p) \\  \mathrm{sin}(\delta_p) \end{array}  \right] 
\end{eqnarray}
the integral in equation (\ref{eq:dens_post3}) can be solved to yield
\begin{eqnarray}
\label{eq:dens_post4}
{\cal P}(\{s_i\}|\{z_p\},\{\theta_n\}) &=&  \sum_{\{N_i\}} {\cal P}(\{s_i\}|\{N_i\})\,{\cal P}(\{N_i\}|\{\vec{y}_p\}) \, .
\end{eqnarray}
Also note that the probability distribution \({\cal P}(\{N_i\}|\{\vec{y}_p\})\) can be written as
\begin{eqnarray}
\label{eq:count_galaxies}
{\cal P}(\{N_i\}|\{\vec{y}_p\}) &=&  \prod_i \delta^K \left(N_i - \sum_p \mathrm{W}(\vec{x}_i -\vec{y}_p) \right) \, ,
\end{eqnarray}
where \(\delta^K(x)\) is the Kronecker delta, \(\vec{x}_i\) is the position of the \(i\)th volume and \(\mathrm{W}(\vec{x})\) is the counting kernel
\begin{equation}
\mathrm{W}(\vec{x})= \mathrm{K}(x^1)\,\mathrm{K}(x^2)\,\mathrm{K}(x^3)\,
\end{equation} 
with \(x^i\) being the components of the Cartesian vector \(\vec{x}\) and $\mathrm{K}(x)$ being
\begin{equation}
\label{equation:kernel_function}
\mathrm{K}(x) = \left \{\begin{array}{c} 1 \, \, \, \,  \mbox{			for \(\left |\frac{x}{\Delta x}\right |<1\)} \\ 0 \, \, \, \,  \mbox{			else} \end{array}\right .
\end{equation} 
and \(\Delta x\) is the length of the volume element.
Therefore, introducing the galaxy number counts
\begin{equation}
\label{eq:appendix_calc_ncounts}
N_i(\{z_p\},\{\theta_p\}) = \sum_p \mathrm{W}(\vec{x}_i -\vec{y}_p)\, ,
\end{equation}
we can solve equation (\ref{eq:dens_post4}) to yield
\begin{eqnarray}
\label{eq:dens_post4}
{\cal P}(\{s_i\}|\{z_p\},\{\theta_n\}) &=&  {\cal P}\left(\{s_i\}|\left \{N_i\left(\{z_p\},\{\theta_p\}\right)\right \} \right) \, .
\end{eqnarray}
Hence, the density posterior distribution can be described in terms of number counts.

\section{redshift posterior distribution}
\label{Appendix_reddist}
In this section we will derive the expression for the redshift posterior distribution of an individual galaxy.
Making frequent use of Bayes rule we can write the conditional redshift posterior distribution described in section \ref{red_post_dist} as
\begin{eqnarray}
\label{eq:cond_redshift_posterior_1a}
{\cal P}(z_n| \{\theta_p \},\{u_p\},\{s_i\},\{z^{obs}_p\}) &=& {\cal P}(z_n) \, \frac{{\cal P}(\{\theta_p \},\{u_p\},\{s_i\},\{z^{obs}_p\}|z_n)}{{\cal P}(\{\theta_p \},\{u_p\},\{s_i\},\{z^{obs}_p\})} \nonumber \\
&=& {\cal P} ( \theta_n ,z_n| \{\xi_p \},\{u_p\} ) \nonumber \\
& & \times \frac{{\cal P}(\{s_i\}|\{\theta_p \},\{z_p\})}{{\cal P}(\{s_i\}|\{\xi_p \},\{u_p\})} \nonumber \\
& & \times \frac{{\cal P}(\{z^{obs}_p\}|\{\theta_p \},\{z_p\},\{s_i\})}{{\cal P}(\{z^{obs}_p\},\Theta_n|\{\xi_p \},\{u_p\},\{s_i\})}\, , \nonumber \\
\end{eqnarray}
We will now consider the probability distribution \({\cal P} ( \theta_n ,z_n| \{\xi_p \},\{u_p\} ) \), which basically describes a random draw of a particle out of an ensemble of particles.  
Changing the coordinates from spherical redshift coordinates to co moving Cartesian coordinates yields
\begin{eqnarray}
\label{eq:1d_cond_redshift_prior_b}
{\cal P} ( \theta_n ,z_n| \{\xi_p \},\{u_p\} )&=& \left | \left .\frac{\partial r(z)}{\partial z}\right |_{z=z_n}\, r(z_n)^2 \, \mathrm{sin}(\delta_n) \right | {\cal P} ( \vec{x}_n| \{\xi_p \},\{u_p\} ) \nonumber \\
\end{eqnarray}
where \(\partial r(z)/\partial z\) is the derivative of the co-moving radial distance with respect to the redshift \(z\).
The Poissonian assumption for the galaxy distribution, assumes, that the number counts, and hence the number of particles inside individual cells, are independent. Further, the Poissonian model assumes the particles inside a volume element to be homogeneously distributed. We therefore yield
\begin{eqnarray}
\label{eq:1d_cond_redshift_prior_b}
{\cal P} ( \vec{x}_n| \{\xi_p \},\{u_p\} ) &=& \sum_{\{M_i\}} {\cal P} ( \{M_i\}| \{\xi_p \},\{u_p\} )\, {\cal P} ( \vec{x}_n| \{M_i\}) \nonumber \\
 &=& \sum_{\{M_i\}} {\cal P} \prod_i \delta^{K}\left( M_i - \sum_p  W\left(\vec{x}_i - \vec{x}(\xi_p,u_p)\right)\right)\, \nonumber \\
 & & \times {\cal P} ( \vec{x}_n| \{M_i\}) \nonumber \\
 &=& {\cal P} ( \vec{x}_n| \{M_i(\{\xi_p \},\{u_p\})\}) \nonumber \\
\end{eqnarray}
In the following we will omit the arguments of the number counts \(\{M_i(\{\xi_p \},\{u_p\})\}=\{M_i\}\). We can then write
\begin{eqnarray}
\label{eq:1d_cond_redshift_prior_b}
{\cal P} ( \vec{x}_n| \{M_i \}) &=& \frac{{\cal P} ( \vec{x}_n, \{M_i \})}{{\cal P} (\{M_i \})} \nonumber \\
&=& \frac{1}{{\cal P} (\{M_i \})} \sum_{\{N_i\}} {\cal P} (\{N_i\} )\, {\cal P} ( \vec{x}_n, \{M_i \}|\{N_i\})\nonumber \\
&=& \frac{1}{{\cal P} (\{M_i \})} \sum_{\{N_i\}} {\cal P} (\{N_i\} )\, {\cal P} ( \vec{x}_n|\{N_i\}) {\cal P} (\{M_i \}|\{N_i\}, \vec{x}_n)\nonumber \\
&=& \frac{1}{{\cal P} (\{M_i \})} \sum_{\{N_i\}} {\cal P} (\{N_i\} )\, {\cal P} ( \vec{x}_n|\{N_i\}) \nonumber \\
& & \times \, \prod_i \delta^K\left(M_i - (N_i - W(\vec{x}_i -\vec{x}_n))\right)\nonumber \\
&=& \frac{{\cal P} (\{M_i+W(\vec{x}_i -\vec{x}_n)\})}{{\cal P} (\{M_i\})} \, {\cal P} ( \vec{x}_n|\{M_i+W(\vec{x}_i -\vec{x}_n)\}) \nonumber \\
\end{eqnarray}
By noting that \(\{M_i+W(\vec{x}_i -\vec{x}_n)\}=\left \{N_i\left(\{z_p\},\{\theta_p\}\right) \right \}= \{N_i \}\) we can write
\begin{eqnarray}
\label{eq:1d_cond_redshift_prior_b}
{\cal P} ( \vec{x}_n| \{M_i \}) &=& \frac{{\cal P} (\{N_i\})}{{\cal P} (\{M_i\})} \, {\cal P} ( \vec{x}_n|\{N_i\}) \, .
\end{eqnarray}
The probability \({\cal P} ( \vec{x}_n|\{N_i\})\) is merely the probability of randomly picking a particle at a specific position conditional on the set of number counts.
Hence we can write
\begin{eqnarray}
\label{eq:1d_cond_redshift_prior_b}
 {\cal P} ( \vec{x}_n|\{N_i\}) &=& \sum_i \frac{W(\vec{x}_i-\vec{x}_n)}{\Delta V}\, \frac{N_i}{N_{tot}} \, ,
\end{eqnarray}
where \(N_{tot}\)  is the total number of particles and \(\Delta V\) is the volume of the volume element.
Next we will consider the fraction
\begin{eqnarray}
\label{eq:1d_cond_redshift_prior_b}
\frac{{\cal P}(\{s_i\}|\{\theta_p \},\{z_p\})}{{\cal P}(\{s_i\}|\{\xi_p \},\{u_p\})} &=& \frac{{\cal P}\left(\{s_i\} \left |\left \{N_i\left(\{z_p\},\{\theta_p\}\right) \right \} \right. \right)} {{\cal P}\left(\{s_i\} \left |\left \{M_i\left(\{u_p\},\{\xi_p\}\right) \right \} \right. \right)} \nonumber \\
&=& \frac{{\cal P}\left(\{s_i\} \left |\left \{N_i \right \} \right. \right)} {{\cal P}\left(\{s_i\} \left |\left \{M_i \right \} \right. \right)} \, ,
\end{eqnarray}
where we used the result, obtained in appendix \ref{Appendix_densitypost}, that the density posterior solely depends through the number counts on the set of galaxy coordinates.
As can be easily seen by applying Bayes law we can write this fraction as the ratio of the Poissonian likelihoods
\begin{eqnarray}
\label{eq:Lh_ratio_1}
\frac{{\cal P}(\{s_i\}|\{\theta_p \},\{z_p\})}{{\cal P}(\{s_i\}|\{\xi_p \},\{u_p\})} &=& \frac{{\cal P}\left(\left \{M_i \right \}\right)}{{\cal P}\left(\left \{N_i \right \}\right)}\frac{{\cal P}\left(\{N_i\} \left |\left \{s_i \right \} \right. \right)} {{\cal P}\left(\{M_i\} \left |\left \{s_i \right \} \right. \right)} \nonumber \\
&=& \frac{{\cal P}\left(\left \{M_i \right \}\right)}{{\cal P}\left(\left \{N_i \right \}\right)} \, \prod_i \frac{M_i!}{N_i!} \, \lambda_i^{N_i-M_i}\, ,
\end{eqnarray} 
with \(\lambda_i=\left(R_i \bar{N}(1+B(s)_i)\right)\) being the intensity of the Poisson process. Since the number densities \(\{N_i\}\) and \(\{M_i\}\) differ only by one at the position \(\vec{x}_n\) of the galaxy under consideration, the ratios of the product in equation (\ref{eq:Lh_ratio_1}) are all equal to one except at this particular galaxy position.
we can therefore write
\begin{eqnarray}
\label{eq:Lh_ratio_2}
\frac{{\cal P}(\{s_i\}|\{\theta_p \},\{z_p\})}{{\cal P}(\{s_i\}|\{\xi_p \},\{u_p\})} &=& \frac{{\cal P}\left(\left \{M_i \right \}\right)}{{\cal P}\left(\left \{N_i \right \}\right)} \, \prod_i \, \left( \frac{\lambda_i}{N_i}\right) ^{W(\vec{x}_i-\vec{x}_n)} \, .
\end{eqnarray} 
Inserting all results into equation (\ref{eq:cond_redshift_posterior_1a}) yields
\begin{eqnarray}
\label{eq:cond_redshift_posterior_1b}
{\cal P}(z_n| \{\theta_p \},\{u_p\},\{s_i\},\{z^{obs}_p\})&=& \left | \left .\frac{\partial r(z)}{\partial z}\right |_{z=z_n}\, r(z_n)^2 \, \mathrm{sin}(\delta_n) \right | \nonumber \\
& & \times \sum_i \frac{W(\vec{x}_i-\vec{x}_n)}{\Delta V}\, \frac{N_i}{N_{tot}} \nonumber \\
& & \times \prod_i \, \left( \frac{\lambda_i}{N_i}\right) ^{W(\vec{x}_i-\vec{x}_n)} \nonumber \\
& & \times \frac{{\cal P}(\{z^{obs}_p\}|\{\theta_p \},\{z_p\},\{s_i\})}{{\cal P}(\{z^{obs}_p\},\Theta_n|\{\xi_p \},\{u_p\},\{s_i\})}\, , \nonumber \\
\end{eqnarray}
Since the product in the third line of equation (\ref{eq:cond_redshift_posterior_1b}) is only different from one at the galaxy position \(\vec{x}_n\) we can write
\begin{eqnarray}
\label{eq:cond_redshift_posterior_1c}
{\cal P}(z_n| \{\theta_p \},\{u_p\},\{s_i\},\{z^{obs}_p\})&=& \left | \left .\frac{\partial r(z)}{\partial z}\right |_{z=z_n}\, r(z_n)^2 \, \mathrm{sin}(\delta_n) \right | \nonumber \\
& & \times \sum_i \frac{W(\vec{x}_i-\vec{x}_n)}{\Delta V}\, \frac{\lambda_i}{N_{tot}} \nonumber \\
& & \times \frac{{\cal P}(\{z^{obs}_p\}|\{\theta_p \},\{z_p\},\{s_i\})}{{\cal P}(\{z^{obs}_p\},\Theta_n|\{\xi_p \},\{u_p\},\{s_i\})}\, , \nonumber \\
\end{eqnarray}
Also note, since in this work we will use Metropolis-Hastings algorithms to explore the conditional redshift posterior distribution, we do not require to explicitly calculate the normalization constant of the probability distribution given in equation (\ref{eq:cond_redshift_posterior_1c}). Up to an overall normalization constant we can therefore write
\begin{eqnarray}
\label{eq:cond_redshift_posterior_1d}
{\cal P}(z_n| \{\theta_p \},\{u_p\},\{s_i\},\{z^{obs}_p\})&\propto& \left | \left .\frac{\partial r(z)}{\partial z}\right |_{z=z_n}\, r(z_n)^2 \right | \nonumber \\
& & \times\, \sum_i W(\vec{x}_i-\vec{x}_n)\, \lambda_i \nonumber \\
& & \times\, {\cal P}(\{z^{obs}_p\}|\{\theta_p \},\{z_p\},\{s_i\})\, , \nonumber \\
&=& \left | \left .\frac{\partial r(z)}{\partial z}\right |_{z=z_n}\, r(z_n)^2 \right | \nonumber \\
& & \times\, \sum_i W(\vec{x}_i-\vec{x}_n)\, \left(R_i \bar{N}(1+B(s)_i)\right) \nonumber \\
& & \times\, {\cal P}(\{z^{obs}_p\}|\{\theta_p \},\{z_p\},\{s_i\})\, . \nonumber \\
\end{eqnarray}
This final result demonstrates that the conditional redshift posterior is essentially proportional to the intensity function of the Poissonian likelihood. It is also interesting to note, that the right hand side of equation (\ref{eq:cond_redshift_posterior_1d}) does not depend on the galaxy coordinates \(\{u_p\}\) and \(\{\xi_p\}\) of all galaxies. This shows, that given the underlying density field \(\{s_i\}\) the individual redshift posterior distribution is conditionally independent of all other galaxies
\begin{eqnarray}
\label{eq:cond_redshift_posterior_1e}
{\cal P}(z_n| \{\theta_p \},\{u_p\},\{s_i\},\{z^{obs}_p\}) = {\cal P}(z_n|\{s_i\},\{z^{obs}_p\}) \, .
\end{eqnarray}
If we make the additional assumption, that the statistical process by which the observed redshifts \(z^{obs}_p\) for individual galaxies is independent of the observed redshifts of all other galaxies we may write the redshift likelihood as
\begin{eqnarray}
\label{eq:cond_redshift_posterior_1d}
{\cal P}(\{z^{obs}_p\}|\{\theta_p \},\{z_p\},\{s_i\}) &=& \prod_p {\cal P}(z^{obs}_p|\theta_p ,z_p ,\{s_i\})\, .
\end{eqnarray} 
Using this definition in equation (\ref{eq:cond_redshift_posterior_1d}) and omitting all normalization constants yields
\begin{eqnarray}
\label{eq:cond_redshift_posterior_1d}
{\cal P}(z_n| \{\theta_p \},\{u_p\},\{s_i\},\{z^{obs}_p\})&\propto & \left | \left .\frac{\partial r(z)}{\partial z}\right |_{z=z_n}\, r(z_n)^2 \right | \nonumber \\
& & \times\, \sum_i W(\vec{x}_i-\vec{x}_n)\, \left(R_i \bar{N}(1+B(s)_i)\right) \nonumber \\
& & \times\, {\cal P}(z^{obs}_n|\theta_n , z_n ,\{s_i\})\, . \nonumber \\
\end{eqnarray}
This result permits us to sample each galaxy redshift \(z_n\) individually.
\bsp

\label{lastpage}

\end{document}